%% file: main.tex
\documentclass{xarticle}

\input{preamble}

\begin{document}

\title{\bt: Control Time, Not Flows }
\note{Preprint}

\def\MODE{1}

\if\MODE1\def\makeauthor{\author{%
    Martijn Bastiaan\footnotesymbol{1} 
    \quad
    Christiaan Baaij\footnotesymbol{1} 
    \quad
    Martin Izzard\footnotesymbol{2} 
    \quad\\
    Felix Klein\footnotesymbol{1} 
    \quad
    Sanjay Lall\footnotesymbol{2,3} 
    \quad
    Tammo Spalink\footnotesymbol{2}}
}\fi

\makeauthor

\maketitle

\def\makefeet{\makefootnote{1}{QBayLogic, The Netherlands}
  \makefootnote{2}{Google DeepMind.}
  \makefootnote{3}{Department of Electrical
    Engineering, Stanford University, Stanford, CA 94305. \texttt{lall@stanford.edu}}
}

\if\MODE1\makefeet\fi

\begin{abstract}
This paper presents the first hardware implementation of \bt, a 
decentralized clock synchronization mechanism for achieving logical 
synchrony in distributed systems. We detail the design and 
implementation of an 8-node \bt network using off-the-shelf FPGA 
boards and adjustable clock sources. Through experiments with various 
network topologies, including fully connected, hourglass, and cube, 
we demonstrate the effectiveness of \bt in aligning node 
frequencies and bounding buffer excursions. We collect and analyze 
frequency, buffer occupancy, and logical latency data, validating 
the hardware's performance against theoretical predictions and 
simulations. Our results show that \bt achieves tight frequency 
alignment, robustly handles varying physical latencies, and 
establishes a consistent notion of logical time across the network, 
enabling predictable distributed computation at scale with zero in-band
overhead.
\end{abstract}

\section{Introduction}

There are many applications in modern datacenters which are
implemented effectively using distributed systems. Some applications
benefit from hard guarantees about the state at distinct nodes, such
as databases~\cite{cockroachdb,li_sundial_2020,corbett_spanner_2013}
and financial exchanges~\cite{gupta}, where the system must ensure that all nodes
can reason correctly about the order of transactions. Another important feature in
applications is predictable latency and resource requirements,
important in robotics~\cite{bateni} and in large-scale numerical
computations such as machine-learning training and
inference~\cite{cowan}.

The recently proposed approach of \emph{logical synchrony}~\cite{ls}
offers a new direction for mitigating the complexities of such
distributed computing applications. In a logically synchronous system,
data may be transmitted between nodes without requiring backpressure
mechanisms or flow control. Retransmissions are handled at the
application level, so that at the lower levels data transmission is
predictable. For two directly-connected nodes, the sender can count
frames sent and the receiver can count frames received, and since there
are no retransmissions these counters are sufficient for the receiver
to identify particular frames from the sender. As discussed
in~\cite{ls}, these counters allow joint
ahead-of-time scheduling of compute and communications.

Without retransmissions or backpressure, clock synchronization becomes the
fundamental mechanism which enables the network to function. If, on
average, a sender is transmitting data faster than the receiver is
processing it, then the receiver's buffer will overflow, and
conversely if the sender is transmitting too slowly, then the buffer
will underflow and the receiver will be starved. These situations are
prevented by ensuring that the clocks at all nodes have, averaged over
time, the same frequency. This property, known as \emph{syntony}, is
much weaker than synchronization of absolute time, as provided by 
PTP~\cite{ptp} for example. The \bt
mechanism~\cite{spalink_2006,bms} offers a zero-overhead method for
achieving syntony.  It is the purpose of this paper to implement this
mechanism in hardware, and demonstrate that such networks can be
practically built, and data can be sent between nodes without the need
for flow control.

\begin{figure}[htb]
  \centering
  \medskip
  \hbox to \linewidth{\hss\begin{overpic}[abs,width=80mm,unit=2mm]{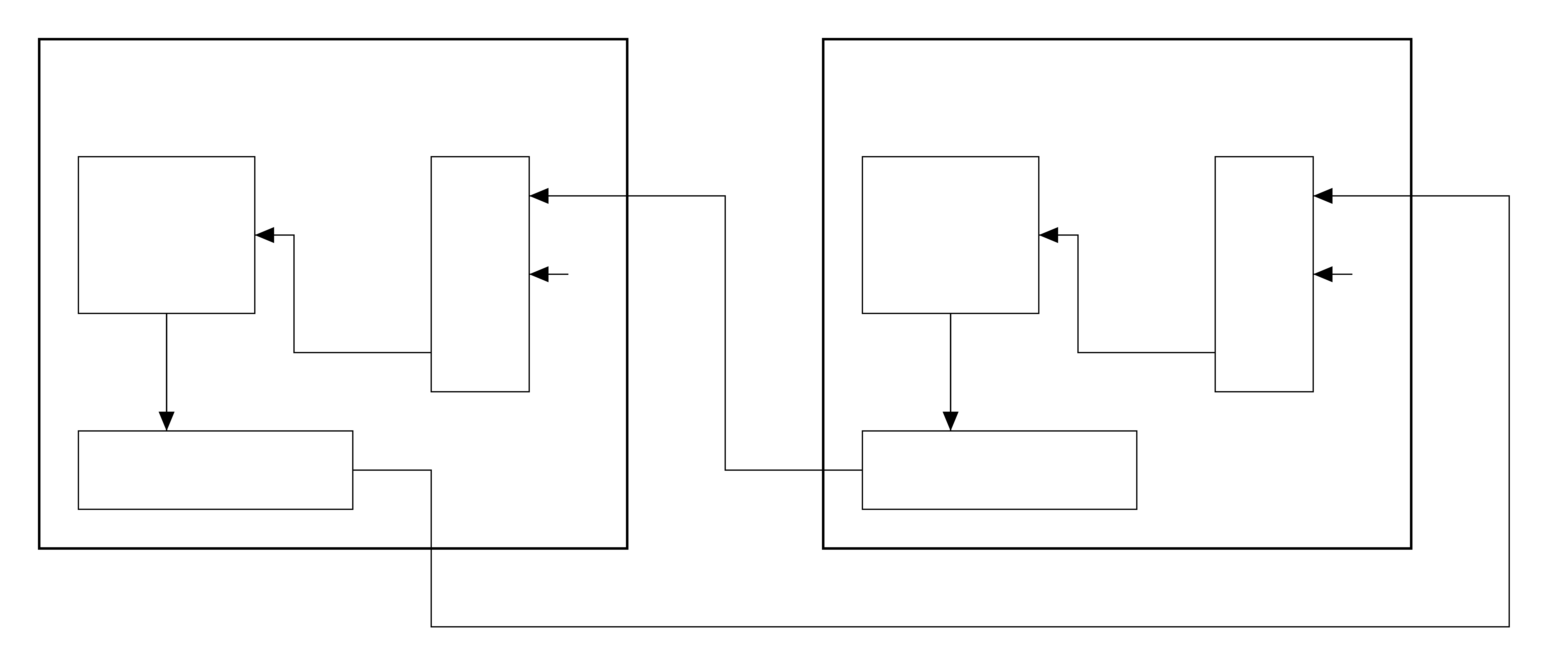}
      \put(5.5,4.8){\clap{\fivept send memory}}
      \put(4.25, 11){\vcent{\clap{\fivept processor}}}
      \put(12.25, 10){\clap{\vcent{\rotatebox{90}{\fivept receive buffer}}}}
      \put(14.6, 9.8){\colorbox{white}{\fivept midpoint}}
      \put(10.7, 8.2){\llap{\fivept read ptr}}
      \put(14.2, 12.4){\colorbox{white}{\fivept write ptr}}
      \put(2, 15){\scalebox{0.5}{{\doublefont node} $i$}}
      \put(25.5,4.8){\clap{\fivept send memory}}
      \put(24.25, 11){\vcent{\clap{\fivept processor}}}
      \put(32.25, 10){\clap{\vcent{\rotatebox{90}{\fivept receive buffer}}}}
      \put(34.6, 9.8){\colorbox{white}{\fivept midpoint}}
      \put(30.7, 8.2){\llap{\fivept read ptr}}
      \put(34.2, 12.4){\colorbox{white}{\fivept write ptr}}
      \put(22, 15){\scalebox{0.5}{{\doublefont node} $j$}}
      \put(16.5, 1.3){\scalebox{0.5}{{\doublefont link} $i \to j$}}
      \put(17.5, 4.9){\rotatebox{90}{\scalebox{0.5}{{\doublefont link} $j \to i$}}}
    \end{overpic}\hss}
  \caption{Two nodes showing the receive buffers as part of the links.}
  \label{fig:buffers}
\end{figure}
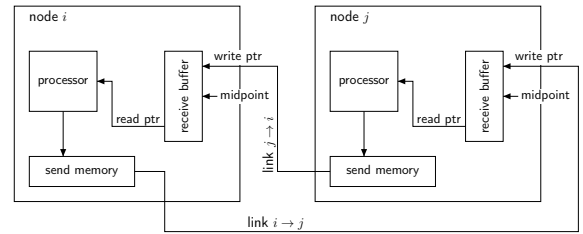

Logical synchrony is still under development. 
There are important issues which have not yet been
resolved, such as failure handling, and these must be addressed in
order to make the system practical. There are also limitations of this
approach; ahead-of-time scheduling may only be used in applications
where communication, computation and resource usage are predictable.
There are also many potential positive consequences of a \bt
system. Simultaneous time-division multiplexing of both compute cycles
and communications becomes possible. Multi-hop communications and
network routing can be ahead-of-time scheduled. Frame arrival and
departure events become logically tied together, allowing precise
coordination and reasoning about the ordering of
events~\cite{lamport}. These topics have been analyzed in~\cite{spalink_2006,bms,ls}
and simulated using the tools called Callisto~\cite{callisto} and
Aegir~\cite{aegir}.  Both the questions raised by these limitations
and the implementation requirements of these higher level mechanisms
are important but beyond the scope of this paper and are not addressed
here.  Instead, we take the important first step of building a
prototype, to show that in principle the ideas of \bt and logical
synchrony are sound and implementable and that the required level of
syntony is achievable in hardware. 

\subsection{Objectives and contributions}

Specifically this paper marks the first hardware realization of a \bt system.
The hardware is open-source and is available at~\cite{qbay}.
We focus on validating the practical feasibility and performance of 
the proposed clock control mechanism in a real-world setting. 
By implementing \bt on an 8-node network of FPGA boards, we demonstrate 
its effectiveness in achieving network-wide frequency alignment and 
bounded buffer excursions under different network topologies.

\subsection{The \bt network}

Each node has a processing unit together with incoming and outgoing
serial data links.  At a node, each incoming data link is connected
via a serdes (serializer-deserializer) to a FIFO, called the
\emph{elastic buffer}, with one such FIFO per link. Data on the serial
links is divided into fixed length frames. On an incoming interface,
as each frame arrives, it is added to the tail of its corresponding
elastic buffer.  In addition, each node has a local clock, whose
frequency is adjustable.  This clock is multiplied to drive the
outgoing serial links.  With every tick of this clock a frame is
removed from the head of the elastic buffer, and moved into memory set
aside for receiving data at the node. Within the same clock period on
each outgoing link a new frame is sent; the data for these frames are
taken from a memory buffer. A two node network is illustrated in
Figure~\ref{fig:buffers}. This structure generalizes to more than two
nodes by allowing multiple incoming and outgoing links per node.

\subsection{Logical synchrony}

The principle of logical synchrony is that \emph{the arrival time of a
frame can be predicted precisely from the departure time.} The meaning
of the word \emph{time} in this statement must be clarified, however.
It does not refer to wall-clock time, or any global notion of time.
Instead each node $i$ has its own clock, which drives the processor,
all of the outgoing serdes, and a counter~$\theta_i$ which therefore
counts outgoing frames. For a frame sent from node $i$ to node $j$,
the departure time is the value of $\theta_i$ at the time it is sent,
and the arrival time is that value of $\theta_j$ at the time it is
removed from the FIFO at node~$j$. This is called \emph{logical}
synchrony, since the times referred to are integers, and they are
exactly defined; there is no requirement for error-bounds on their
values.

We refer to the counter values as \emph{localticks}.  They are the
unit of time at a node; each node has its own unit of time.  The \bt
mechanism, discussed below, ensures that all of the localticks have
approximately the same frequency, on average, but these frequencies
vary as the clocks are continuously adjusted. There is no global clock
in the system. Even though the clocks are ticking at roughly the same
frequency, there is no mechanism which sets the counters to any common
value, and so the counter value (\ie, the absolute time measured
in localticks) is not synchronized across the network and may differ
greatly from one node to another.

Suppose at localtick~$n_i$ node~$i$ sends a frame to node~$j$.  Upon 
arrival, node~$j$ inserts this frame into the tail of its elastic buffer. 
At a later localtick~$n_j$, node~$j$ pops the frame from the head of 
its buffer and passes that frame to the core at node $j$, indicating 
the frame has been \emph{received} by node~$j$.  The next frame sent 
from node~$i$ is sent at localtick~$n_i + 1$, and is is received at 
localtick~$n_j+1$, since frames are received in the same order they are 
sent. As a result, the difference between the arrival time at node~$j$ 
and the transmission time at node~$i$ is a 
\emph{constant}~$\lambda_\itoj = n_j - n_i$, called the 
\emph{logical latency}.  Notice that the transmission time is measured
in localticks at node $i$ and the arrival time is measured in
different units, specifically the localticks at node $j$. 

\subsection{Consequences of logical synchrony}

The fact that logical latency is constant means that data transmission
and arrival times are \emph{schedulable} ahead of time, before any
code is executed.  From the perspective of applications running on the
system, all they need to know about the network is the logical
latencies.  Such a network is represented by a graph of nodes with
directed edges corresponding to links and with a logical latency
associated with each edge.  This graph is called a \emph{logical
synchrony network}~\cite{ls}. It holds sufficient information to
schedule distributed computation. Moreover, the predictable
cycle-accurate timing of all communications means that one can
implement distributed computation while avoiding complex mechanisms
such as barriers.

Logical synchrony~\cite{ls} is therefore a computation and
communication framework for describing distributed computation
networks, which neither requires asynchronous handshakes nor the
distribution of wall clock time for offering synchronous network
operation. Instead, the constancy of logical latency is leveraged, and
frames that are exchanged between the processes in the network are
related only via their causal order. It is shown in~\cite{ls} that the
model always leads to an acyclic execution graph as long as the
utilized FIFO buffers do not over- or underflow.  This allows the use
of the distributed event framework of Lamport~\cite{lamport}.

A practical consequence of logical synchrony as implemented by \bt is
that we can treat multiple, independently clocked nodes as
related. That is, the difference between one node's clock counter and
a clock counter received from another node is constant. This is
similar to crossing two related clock domains on a single chip, where
one can ignore FIFO backpressure signals between the two domains and
therefore never skip a cycle (In practice, crossings between related
domains can use simpler hardware than full FIFOs.) A \bt system lifts
this property to a distributed system, opening the door to a wide
range of applications. For example, one might use \bt to build very
deep computation pipelines, with feedback loops, and these would
be logically synchronized even across multiple data centers.

\subsection{The \bt mechanism}

We give a brief overview of the \bt mechanism here. Further details
and background is given in~\cite{bms,ls}.
At any node, the local clock frequency determines the frame transmission 
rate, while the neighboring node clock frequencies determine the frame 
arrival rate. Therefore, the elastic buffer occupancy will fluctuate. 
If the clock at node~$i$ is faster than that at node~$j$,
then the elastic buffer at node~$i$ for the link~$j \to i$ will start
to drain. Conversely, it will start to fill if the relative speeds are
reversed. Therefore, the occupancy of the elastic buffers provides a
signal informing the node about the relative frequency of the node
with respect to its neighbors.
With only two nodes, there is a simple mechanism for controlling
frequency. Each node has an elastic buffer. When its elastic buffer
starts to drain, it should reduce the frequency of its clock, and when
its elastic buffer starts to fill, it should increase the frequency.

The resulting system behavior is analyzed in~\cite{bms,reframing}. The
fundamental property obtained is that the system is \emph{stable}, in that
all buffer occupancies and all frequencies converge to a steady
state value. Furthermore, all frequencies become \emph{aligned} over
time; that is, they converge to the same value.  The key property that
this mechanism provides therefore is that it prevents the elastic
buffers from overflowing or underflowing, and it does so by bringing
all frequencies to a common value.

\subsection{Overhead}

One aspect of the \bt mechanism has \emph{zero} overhead, and that is
the basic frequency synchronization scheme. This is because, in
physical serial links such as ethernet, bits are being continually
transmitted across the link. The physical wire is always full of ones
and zeros; sometimes these are application or system data, but the
rest of the time this is simply junk bits sent in order to maintain
frequency lock when the application has no data to send. The frequency
control mechanism continues to operate in the same way, as it is not
affected by the contents of the frames, and does not require that
specific frames are sent at particular times. It simply observes the
frequency of the frame arrival rate.

The frequency is therefore controlled by purely observing the behavior
of the lowest layer of the network. It is transparent to applications
and protocols at higher levels in the network stack,
and it is unaffected by them. The frames serve two purposes; first
they (sometimes) contain data, which applications at the nodes send to
each other, and second the rate of frame arrival and departure
provides timing information. This timing information has no overhead
in terms of data, since no explicit communication is used to transmit
the timing signal.  In this sense, the \bt mechanism incurs no in-band
signaling overhead.

The oscillators at each node are adjusted by a feedback control
mechanism.  This works to track and compensate for small amounts of
drift in the frequency of the underlying oscillator.  The use of
feedback means that the system is regulated against other physical
variations in the system, such as temperature variation, or variation
in the material or other physical properties of the links.

\subsection{Comparison with global approaches}

There exist several well-known methods for synchronizing clocks across
a network, include Sonet~\cite{sonet}, SyncE~\cite{syncE}, 
PTP~\cite{ptp}, and White Rabbit~\cite{whiterabbit}.
These systems provide accurate bounds on the difference between the
value of local clocks and global clocks, and can be used to implement
robust higher-level coordination systems for distributed computation.
There are advantages and disadvantages found when comparing this
approach to \bt. A global clock is easily interpretable by
applications, and the ability to synchronize it with wall-clock time
is important in many applications. In \bt, one does not have a global
clock, nor wall-clock time, but the logical time is discrete and
absolute, with no need for error-bounds, and so inter-node
coordination algorithms can be very simple. Since \bt is a very
low-level mechanism with no in-band signaling, coordination can be
performed efficiently at the granularity of clock-cycles.

In conventional networks, errors in data transmission may be handled
by the network stack, with retransmission happening transparently to
the application. In \bt, if computation and computation are scheduled
ahead-of-time, then any communication errors must be detected and
handled by the application code.

\section{Control algorithm}

The precise \bt mechanism for general networks is as follows.
Let 
the \emph{unadjusted} frequency of the adjustable clock at node~$i$ be
$\wu_i$.  We choose a multiplier~$1 + \crel_i$ such that the clock
will have frequency~$\omega_i$ with
\[
\omega_i(t) = \wu_i ( 1 + \crel_i(t))
\]
where $c^\text{rel}_i$ is called the \emph{relative correction}
at node~$i$. Denote by $\beta_\jtoi$ the buffer occupancy
of the elastic buffer at node $i$ for the link $j \to i$.
Periodically, at each node $i$ we observe the occupancy
of all of its elastic buffers. Then we choose the relative correction according to
\begin{equation}
  \crel_i(t) = k_p \sum_{j \mid \jtoi} (\beta_\jtoi - \beta^\text{off})
  \label{eqn:pcontroller}
\end{equation}
In this equation the sum is over all nodes $j$ for which there is a
link from $j$ to $i$.  That is, we set the relative frequency of the
clock in proportion to the the sum of the elastic buffer occupancies.
Here, we normalize the buffer occupancy~$\beta_\jtoi$ by subtracting
the offset~$\beta^\text{off}$, which is chosen to be the midpoint of
the elastic buffer memory. By doing this we ensure that the elastic buffers
have room to both shrink and grow as necessary to control the frequencies.

The positive constant $k_p$ is called the \emph{gain}.  Intuitively,
this does the right thing: frequency gets increased when the buffer
occupancies are large. It is shown in~\cite{bms,reframing,mbo} that if
$k_p$ is sufficiently small then the system is \emph{stable}.

\section{Hardware components}

Each \bt node consists of three boards; an FPGA development board, a
mezzanine card with additional network interfaces, and a variable
clock source board. Specifically, these are the Kintex Ultrascale FPGA
on a KCU105 evaluation kit, the TEF0008 FPGA mezzanine card (FMC), and
the \mbox{SI5395J-A} adaptable clock source on a Skyworks
\mbox{SI5395J-A-EVB} evaluation board.

The eight nodes are interconnected via the two dedicated small
form-factor pluggable (SFP) connectors of the KCU105 evaluation
boards, four additional SFP connections on the TEF0008 FMC card, and a
final copper link via the gigabit transceiver (GTH) TX and RX
subminiature (SMA) differential pair connectors of the board, creating
28 bidirectional links in total.

\begin{figure}[htb]
  \includegraphics[width=0.47\textwidth]{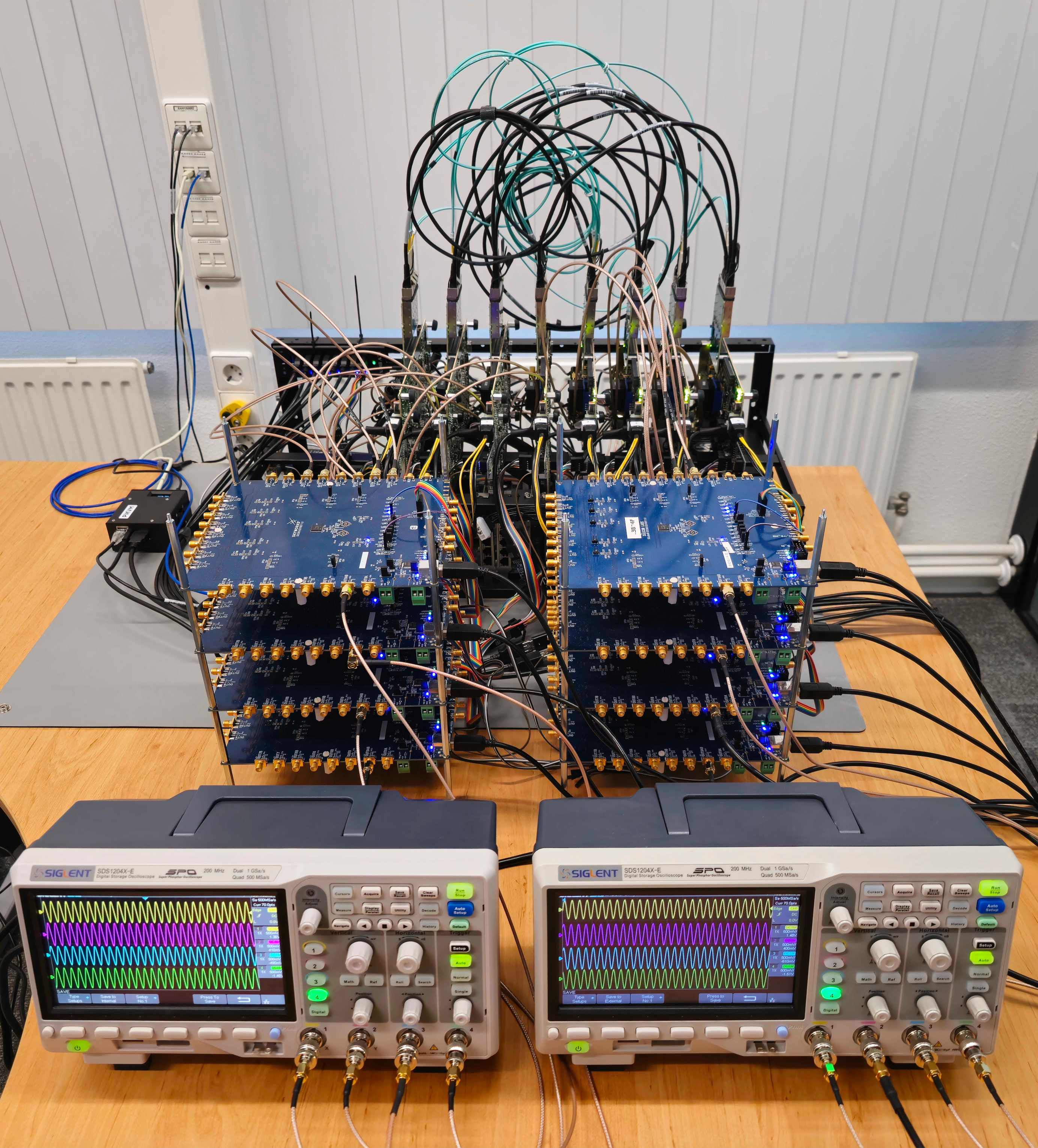}
\centering
  \caption{Physical \bt hardware setup. In front are oscilloscopes, showing (synchronized) clock signals. Behind them are the clock synthesizer boards. In the back are the FPGA development boards.}
  \label{fig:setup}
\end{figure}

The setup is shown in Figure~\ref{fig:setup}. The green boards
in the background are the KCU105 FPGA evaluation kits with the FMC
cards at the top, while the blue boards in the foreground are the
Skyworks SI5395J-A-EVB. Each of the adaptable clock sources drives one
of the FPGAs.

\subsection{Adjustable clock sources}

We use the standard term \emph{parts per million} (ppm) for relative
differences between frequencies, so that for frequencies $\omega_a$ and
$\omega_b$ we say that $\omega_a$ is $\alpha$ ppm higher frequency
than $\omega_b$ if $\omega_a = (1 + \alpha/10^6) \omega_b$. Similarly for
parts per billion (ppb) and parts per trillion (ppt).

The oscillators on the Skyworks clock boards are specified to have an
initial accuracy of \qty{\pm8}{\ppm} when in standard environmental
conditions.  Therefore, any two oscillators can differ by
roughly
\qty{16}{\ppm}.  The oscillators are not temperature compensated, and
the specification lists a drift of \qty{\pm88}{\ppm} over a
temperature range from \qty{-40}{\celsius} to
\qty{125}{\celsius}. Finally, all other environmental factors are
bound within \qty{\pm2}{\ppm}, for an overall maximum deviation of
\qty{\pm98}{\ppm}. It is worth noting that without the frequency
control provided by the \bt mechanism, the buffer occupancies will
rapidly over- or underflow.

The oscillators are paired with a frequency multiplier/divider circuit
that can scale the frequency from \qty{100}{\hertz} to
\qty{700}{\mega\hertz} while the clocks are active. The frequencies
can be adjusted in steps of \qty{1}{ppt}, or an
arbitrary multiple of it. We configure the boards to have a step size
of \qty{0.01}{\ppm}. The frequency can be changed at intervals of
\qty{1}{\micro\second}.  A frequency increase (\finc) or decrease
(\fdec) is requested over dedicated pins.

A node with $m$ incoming links will have $m+3$ clock domains; these
are the incoming link clocks, the node clock, the outgoing link clock,
and a system clock for programming the boards.  Each node local clock
is set at nominal frequency 125MHz.  This clock is multiplied by a
factor of 80 to drive the outgoing serial links at 10GHz.  Each frame
corresponds to 80 transmitted bits.  We use 8b/10b encoding, so every
8 bits of data are encoded into 10 bits on the line.  Each frame
contains 64 bits of useful data.  Each node has~7 incoming links and~7
outgoing links. Some use copper as the transmission medium, others use
fiber.

\section{Hardware design}
We use Clash as the primary hardware description language. Clash is a freely distributed, OSS back-end for the \emph{Glasgow Haskell Compiler} emitting Verilog/VHDL. It allows us to describe hardware using Haskell's strong type system and abstraction capabilities. For this design, three features of Clash were particularly useful:

\begin{itemize}

\item \emph{Clock domain crossing safety}: Clash's type system encodes
  clock domains, preventing implicit (or accidental) crossings. In
  other hardware design languages, this is a common source of bugs
  typically only caught after synthesis. This was mostly relevant for
  the design of the transceiver logic
  
\item \emph{Auto-pipelining floating point operations}: floating point operations take multiple clock cycles to compute. Inputs and intermediate results therefore need to be appropriately delayed. For example, in the equation $ a \cdot b + c $, $c$ needs to be delayed as long as it takes to compute multiplication. Clash can keep track of this on the type level, allowing for functions that insert the appropriate number of registers automatically. See \Cref{fig:code-snippet} for part of the implementation of clock control.
\item \emph{General purpose programming}: Not only can we describe hardware with Clash, we can also use Haskell's ecosystem to generate experiments, process the captured data, simulate the design, and generate diagrams. Any types or functions used in hardware are naturally in sync with the rest of the tooling due to the shared language front-end.
\end{itemize}

\begin{figure}[htb]
  \begin{quote}
    \fontsize{8pt}{11}\selectfont
    \input{default.pygstyle}
    \input{listing1.pygtex}
  \end{quote}

\caption{Part of the clock control algorithm. Note the use of \texttt{delayI}, which automatically inserts the appropriate number of registers to align the signals.}
\label{fig:code-snippet}
\end{figure}

\begin{figure}[htb]
   \centering
   \includegraphics[width=0.9\linewidth]{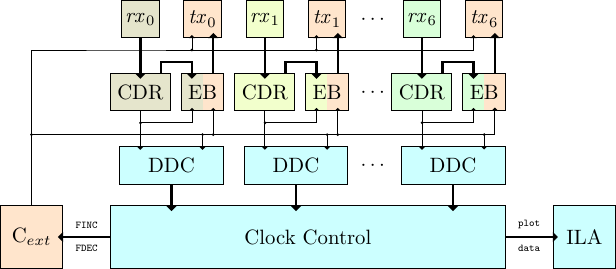}
   \vspace*{8mm}
    \caption{Block diagram of clock control circuit of a single
      node. Thin lines indicate clock signals, while thick lines carry
      data. Different colors indicate different clock domains. Here
      CDR denotes \emph{clock and data recovery}, EB denotes \emph{elastic buffer},
      and DDC blocks are \emph{domain difference counting}. The ILA is the \emph{integrated logic
      analyzer}, and $\textup{C}_\textit{ext}$ the adjustable clock board.}
    \label{fig:cc-block-diagram}
\end{figure}

\subsection{System bootup}
\label{subsection:boot}

The different components of our hardware setup need to be turned on in the correct order to bring the whole system into the desired logically synchronous state. The particular boot process can be summarized as follows:

\begin{enumerate}

\item As the first step, the whole network is powered including all the clock boards and the FPGAs.

\item Next, each clock boards is programmed by the corresponding connected FPGA via SPI to output a 200Mhz clock (according to internal reference of the SI5395J-A). This includes the step size of the \finc~/~\fdec pins is set as required.

\item Then, the transceiver links get initiated including the handshakes for frame alignment and clock recovery. Pseudo-random data is used to negotiate the connections at this point. In case a connection cannot be negotiated on the first attempt, the procedure gets repeated until it succeeds. All transceivers must be connected reliably for at least 500ms to continue with the next phase of the boot.

\item Finally, a shared trigger of our external monitoring system initializes clock control and the elastic buffers of all nodes. We start all of the nodes at the same point in time to observe each node's clock behavior, similar to~\cite{bms,ls}. There is no fundamental reason to start all nodes at the same time, but it simplifies the analysis of the system.

\end{enumerate}

\subsection{Domain difference counting}
\label{subsection:ddc}

We use $ \nats_{n} $ to denote the set of unsigned numbers represented
via bit-vectors with $ n $ bits and $ \ints_{n} $ to denote the set of
signed numbers represented by such vectors, with the MSB being the
sign bit. Bit-vectors can be concatenated via the $ \concat $
operator, such that $x \in \nats_{n}$ and $ y \in\nats_{m}$ then
$x \,\concat\, y \in \nats_{n+m} $.

For clock control only the size of the elastic buffer occupancy is
relevant, not the actual data. During the initial clock
synchronization phase, we thus do not need to store any actual data in
the elastic buffers. Instead we can count the number of frames
arriving, $clk_{rx}$, and the number of frames departing, $clk_{tx}$,
and the difference between these two counters is the number of frames
that would be in the elastic buffer.  These counters are called the
\textit{Domain Difference Counters (DDCs)}, and we consider them to
act like \textit{virtual elastic buffers} until all clocks of the \bt
network have been synchronized. The advantage of this approach is
that, during the initial clock synchronization the elastic buffer
occupancies can become large and overflow. Using virtual elastic
buffers avoids this. After synchronization, we can use a
\emph{reframing} procedure~\cite{reframing} to recenter the elastic
buffers.  Data exchange is only of interest when applications are
running, which happens after we have synchronized the network. Hence,
we can make use of the DDCs at first, and then switch to real elastic
buffers afterwards.

\begin{figure}[htb]
  \centering
  \includegraphics{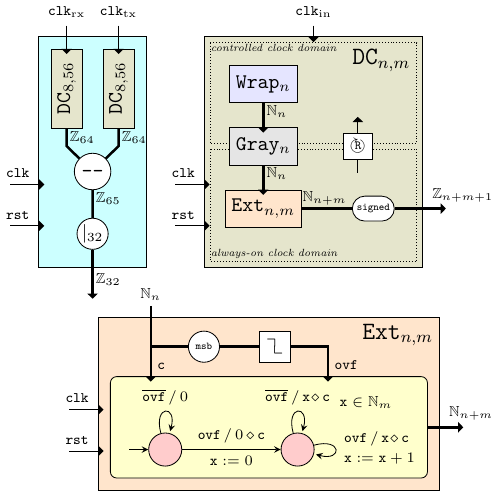}
  \caption{Domain Difference Counting Setup}
  \label{fig:ddc}
\end{figure}

We utilize a virtual buffer size of
$ 2^{32} $. For readability purposes, we map the
occupancy count to a 32-bit signed, where zero indicates \emph{half-full}
(or $2^{31}$ frames). \Cref{fig:ddc} gives an overview of the implementation of the DDCs.

In \Cref{fig:ddc} stateful components have a rectangular shape, while pure
combinational transformations have round shapes. The clock and
reset signals are implicitly shared along components with state unless
separated. $ \texttt{clk} $
and $ \texttt{rst} $ are routed by default. {\syncsym}
denotes a reset synchronizer and {\fdetect} a falling edge
detector. The circuit utilizes
$ \texttt{msb} \from \nats_{n} \to \bools $, extracting the most
significant bit of a bit-vector,
$ \texttt{signed} \from \nats_{n} \to \ints_{n+1} $, to convert an
unsigned number to a signed one via adding the sign bit,
$ |_{n} \from \nats_{n+m} \to \nats_{n} $, for truncating a bit-vector
to size $ n $, and
$ \texttt{--} \from \ints_{n} \times \ints_{n} \to \ints_{n} $, for
computing the difference between two signed numbers. All displayed
stated machines are Mealy machines.

In terms of functionality, the setup utilizes two domain counters
(\texttt{DC}s) are updated independently at the rates
$\texttt{clk}_{\textit{rx}}$ of messages being added to the virtual
 buffer and $\texttt{clk}_{\textit{tx}}$ of messages being removed. Each individual \texttt{DC} consists of a wrapping counter $ \texttt{Wrap}_{n} $ cycling from $ 0 $ up to $ 2^{n}-1 $, a gray counter $ \texttt{Gray}_{n} $, synchronizing the wrapping counter to the always-on clock domain via gray code~\cite{gray}. To prevent circuit logic depth, the gray code synchronization logic is kept fairly small and is later extended to $ 2^{n+m}-1 $.

In our setup, the extended counters are 64 bits. The subtraction's
result is then truncated to 32 bits. This parameter selection is safe,
since it takes a 125 MHz clock about 5 millennia to count to $2^{64}$.
According to the clock generator specifications, clocks can differ
a maximum of $\pm$\qty{98}{\ppm}. In a worst case scenario, it would
take clocks \qty{24}{\hour} of uncorrected use to count to $2^{31}$.
Both of these durations align well within the time it takes to run our
experiments.

\subsection{Clock control}

Our hardware setup requires clock modifications to be executed
stepwise and at discrete points in time via sending \finc~/~\fdec
pulses to the clock boards.  The clock control boards operate in the
following way; if after reset we apply $n^\text{inc}$ pulses to the
\finc pin, and $n^\text{dec}$ pulses to the \fdec pin, then the
resulting oscillator frequency is
\[
\wu (1 + f_s(n^\text{inc} + n^\text{dec}))
\]
Here $f_s$ is the step size.

We must send
the correct number of such pulses to implement
the clock correction.
Instead of returning a continuous
clock correction our controller
returns~$\cinc_i \in \{ -1, 0, 1 \} $ differentiating between
three possible \textit{clock modification directions} at a time, \ie
\begin{equation*}
  \cinc_i(t) = \begin{cases}
    -1 & \text{if } \crel_i(t) < \cest_i(t) \\
        1 & \text{if } \crel_i(t) > \cest_i(t) \\
    0 & \text{otherwise}
  \end{cases}
\end{equation*}
where, $\crel_i$ is the relative correction determined by the clock
control algorithm in equation~\eqref{eqn:pcontroller}.
Here $\cest_i$ is the \textit{estimated frequency correction}
\begin{equation*}
  \cest_i(t) = f_{s} \sum_{t'<t} \cinc_i(t')
\end{equation*}
Note that the sum on the right-hand-side of this equation is a
sum over the history of applied increment/decrement signals applied to
the clock board at all earlier times. The right-hand side of this equation
is therefore equal to the cumulative relative frequency adjustment which
is applied by the clock boards.

The output $\cinc_i$ of the controller triggers pulses at each sample
time $t$ for which $\cinc_i(t) \neq 0 $, sending \finc~/~\fdec pulses
to the clock boards for increasing / decreasing the clock frequency
according to the sign of $\cinc_i$.
When $ c_{d}(t_{i}) = 0 $, the frequency is neither
increased nor decreased. The sampling frequency of the controller is set to the
maximum speed of \qty{1}{\mega\hertz} allowed by the clock boards.

\section{Experiments}

\subsection{Instrumentation}

We periodically store both the clock drift and domain difference counter values
(\Cref{subsection:ddc}). This data is extracted from the FPGAs at the end 
of a run, post processed, and exported as a diagram. For the purpose of 
creating plots, we instrument each of the nodes to measure clock frequency, with an
update frequency of \qty{60}{\milli\second}.  Due to limited
instrumentation capabilities, this leads to visible noise in these
plots. We emphasize that this noise is only in the telemetry, not
something that is present internally within the system.

\subsection{Clock control settings}

\label{subsection:clock-control-settings}

We choose the proportional gain $k_p = 0.25$. This value is picked
experimentally to make the system converge slowly. Choosing a larger
gain will cause faster convergence. The primary limitation here is
that a high gain requires a fast sampling rate to ensure stability
of the feedback system.
Our implementation has sufficiently fast sampling to allow a
substantial higher gain, however the telemetry system is limits our
ability to collect accurate data from the system in that case, and
using a smaller gain allows us to see the system behavior with fewer
measuring artifacts. Settings that let the system converge faster are
discussed in \Cref{subsection:fast-convergence}. After clocks
converge, we enable the elastic buffers for logical latency
transmission. These buffers are 32 elements deep, and are initialized
to half full + 2, \ie, 18 elements.

\subsection{Uniform connectivity --- the fully connected topology}
\label{subsection:fully-connected}
\begin{figure}[htb]
\centering
\includegraphics[width=0.9\linewidth]{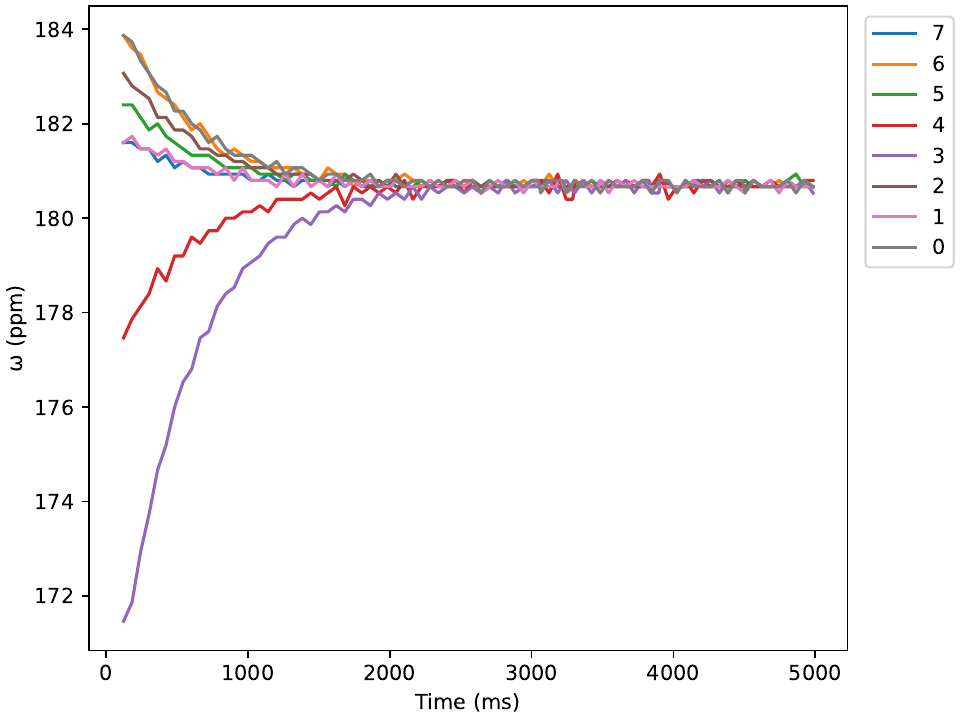}
\caption{Clock frequencies for the fully connected topology}
\label{fig:clocks-2m}
\end{figure}

\begin{figure}[htb]
  \centering
\hskip-10mm\includegraphics[width=0.83\linewidth]{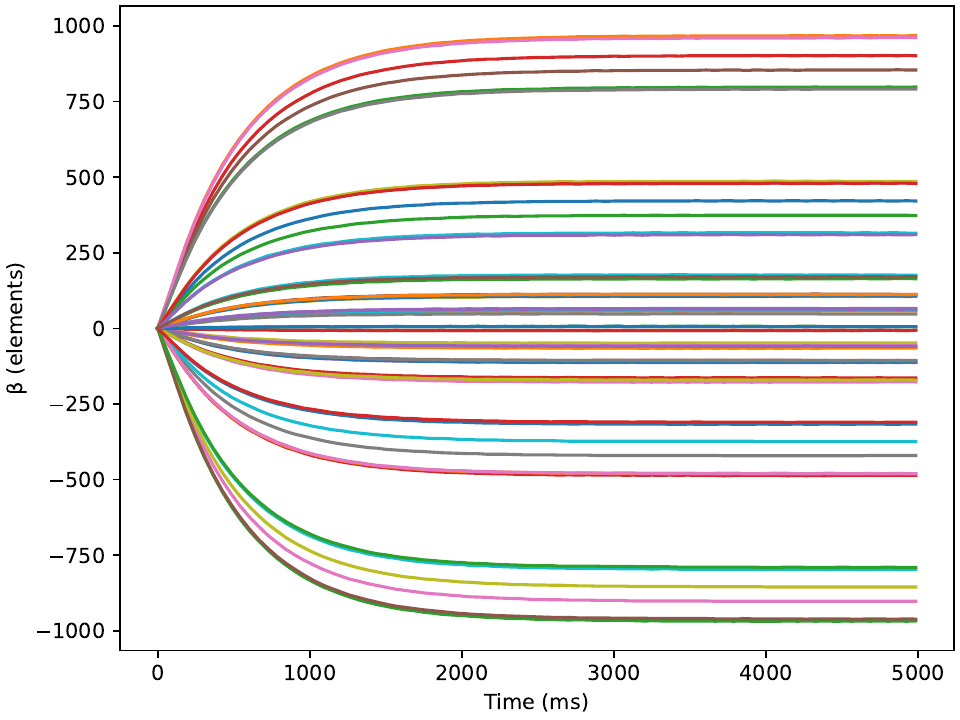}
\caption{Buffer occupancy count for the fully connected topology}
\label{fig:ebs-2m}
\end{figure}

Our first experiment runs on a fully connected network of eight
nodes. Every node is connected to every other node, with
\qty{2}{\meter} of cable or less. The measured clock frequencies for
each node can be found in \Cref{fig:clocks-2m}. The buffer occupancy
count for each node can be found in \Cref{fig:ebs-2m}. The round trip
logical latencies for each node can be found in \Cref{tab:ugns-2m}.

In the clock frequency plot we can see that the frequencies converge
and stay within a \qty{1}{\ppm} range to each other.
Future implementations of \bt's
domain difference counters might operate on clock frequencies native to
their transceivers, allowing for synchronization well within a \qty{1}{\ppb}
margin. Smaller values will allow \bt to use smaller elastic buffers, but
won't affect any other properties.
Note that it is
impossible for the frequencies to be exactly the same, as their
initial offsets are properties of the physical oscillators and are
subject to normal manufacturing variability.  As discussed above, the
clock control mechanism is set so that the frequency is adjustable in
steps of \qty{0.1}{\ppm}. We can therefore interpret the these plots
as evidence that the clock control continuously adjusts the clock
frequencies to be as close as possible to each other, every time under-
and overshooting by a small amount.

In the buffer occupancy plot we observe that each buffer converges to
a stable value. The plot is almost symmetrical, due to the symmetrical
nature of the links and elastic buffers. That is, any node that is
slower than its neighbor will both receive fewer frames (filling its
own buffer) and send fewer frames (emptying the neighbor's buffer).

Finally, the round trip time (RTT) logical latencies all hover around
69. We discuss this in~\Cref{section:long-link}.

\begin{table}[htb]
\centering
\begin{tabular}{c|ccccccc}
  FPGA & $l_1$ & $l_2$ & $l_3$ & $l_4$ & $l_5$ & $l_6$ & $l_7$ \\
  \hline
0    & 67        & 69        & 69        & 68        & 69        & 70        & 68        \\
1    & 69        & 69        & 68        & 69        & 68        & 68        & 68        \\
2    & 69        & 69        & 69        & 69        & 69        & 68        & 67        \\
3    & 67        & 69        & 69        & 69        & 69        & 68        & 67        \\
4    & 68        & 68        & 69        & 69        & 69        & 68        & 68        \\
5    & 68        & 69        & 68        & 68        & 69        & 68        & 68        \\
6    & 68        & 68        & 69        & 69        & 69        & 68        & 67        \\
7    & 68        & 69        & 69        & 69        & 68        & 70        & 67
\end{tabular}
\caption{Round trip logical latencies for the fully connected experiment, where $ l_i $ denotes the $ i $th link of each node according to some fixed index mapping}
\label{tab:ugns-2m}
\end{table}

\subsection{Non-uniform connectivity --- the  hourglass topology}
\begin{figure}[htb]
  \centering
  \hbox to \linewidth{
    \hss\vcent{\includegraphics[width=0.26\linewidth]{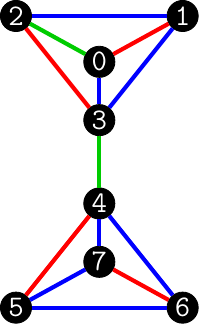}}
    \hss
    \vcent{\includegraphics[width=0.26\linewidth]{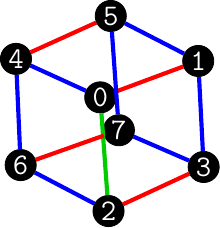}}
    \hss}
  \caption{Hourglass and cube topologies used in the non-uniform
    connectivity experiments. Green links indicate fiber connections,
    blue links Direct Attach (copper) connections and red links copper
    connections using an SMA connector.}
  \label{fig:hourglass}
\end{figure}

\begin{figure}[htb]
  \centering
  \includegraphics[width=0.9\linewidth]{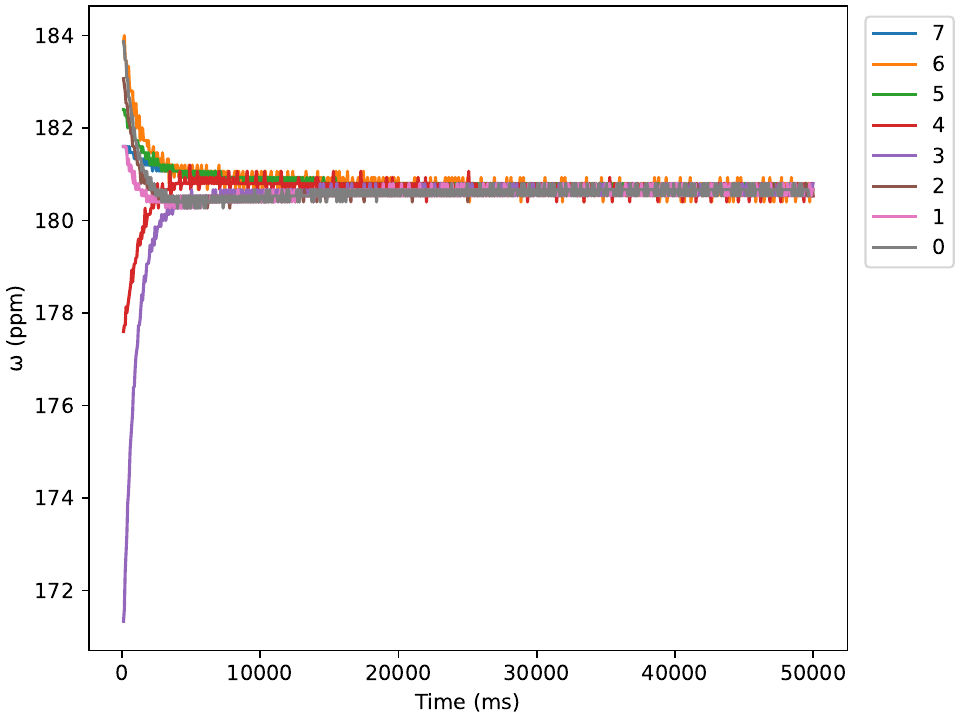}
  \caption{Clock frequencies for the non-uniform, hourglass connectivity experiment}
  \label{fig:clocks-hourglass-2m}
\end{figure}

\begin{figure}[htb]
  \centering
\hskip -10mm\includegraphics[width=0.82\linewidth]{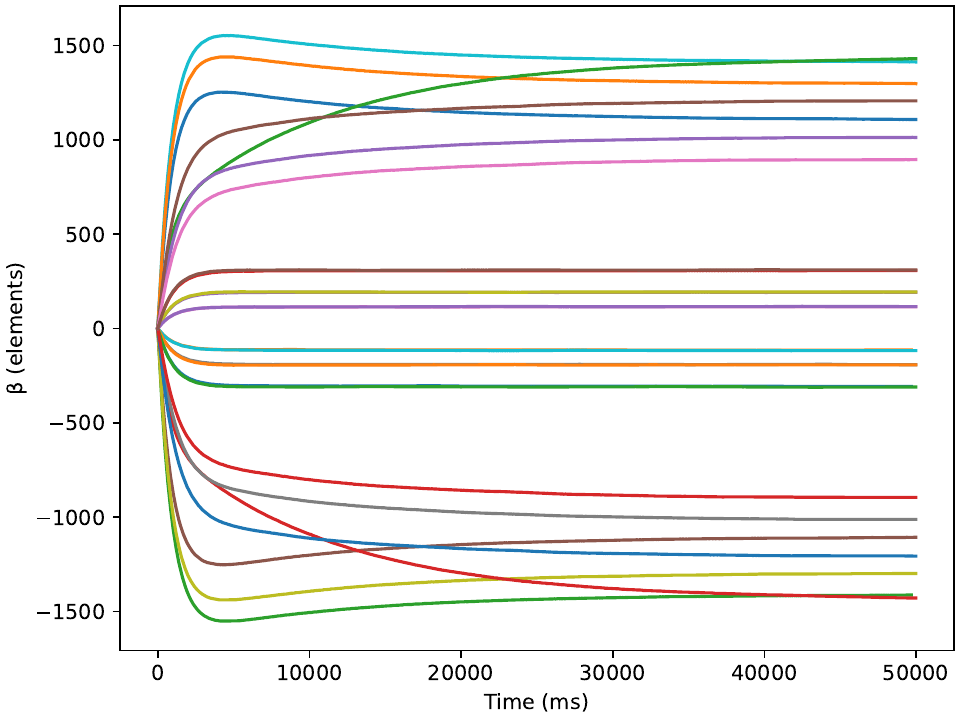}
\caption{Buffer occupancy count for the non-uniform, hourglass topology}
\label{fig:ebs-hourglass-2m}
\end{figure}

For our second experiment, we use the hourglass topology shown in \Cref{fig:hourglass}. The clock frequencies for each node can be found in \Cref{fig:clocks-hourglass-2m}. The buffer occupancy count for each node can be found in \Cref{fig:ebs-hourglass-2m}.

The hourglass topology differs from the fully connected topology in that the nodes are not all connected to each other. Instead, two fully connected subgraphs of four nodes each are connected by a single link. Because clock control does not prefer one neighbor to the other, we expect the nodes in the subgraphs to converge to a similar frequency sooner than two nodes in different subgraphs. This is indeed what we observe in the clock frequency plot. The most notable example of this behavior can be seen for node 4 (red), which first gets pulled up to the other nodes in its groups (nodes 5, 6, and 7), after which it gets pulled down again by the nodes in the other group (nodes 0, 1, 2, and 3). Similar behavior can be observed in the buffer occupancy plot.

\subsection{Non-uniform connectivity --- the cube topology}
\begin{figure}[htb]
  \centering
  \includegraphics[width=0.9\linewidth]{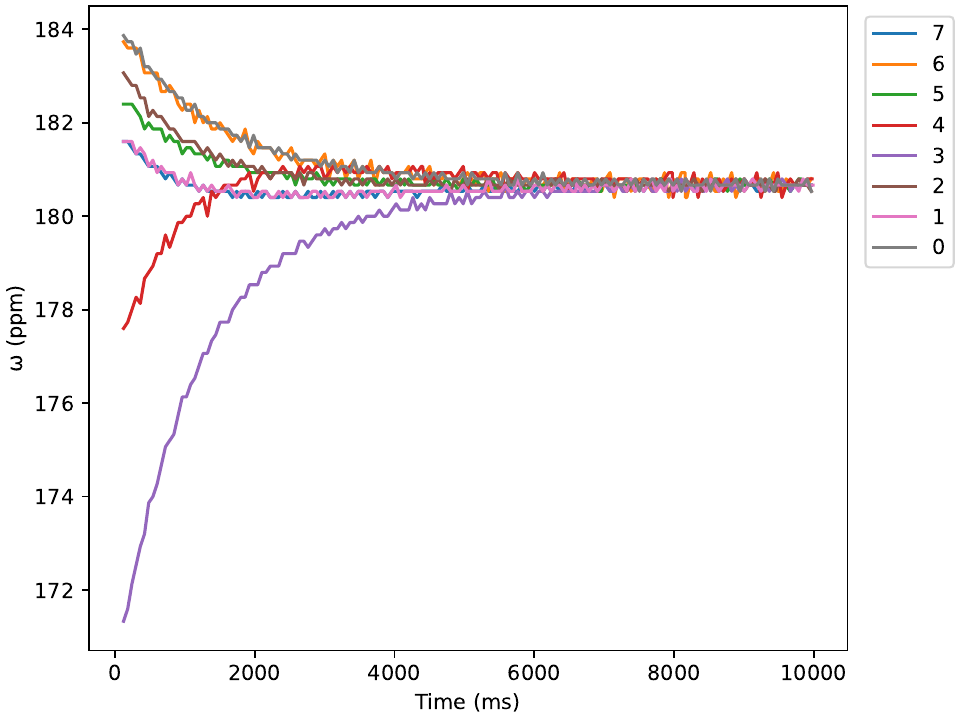}
  \caption{Clock frequencies for the non-uniform, cube topology}
  \label{fig:clocks-cube-2m}
\end{figure}

\begin{figure}[htb]
  \centering
\hskip-11mm\includegraphics[width=0.82\linewidth]{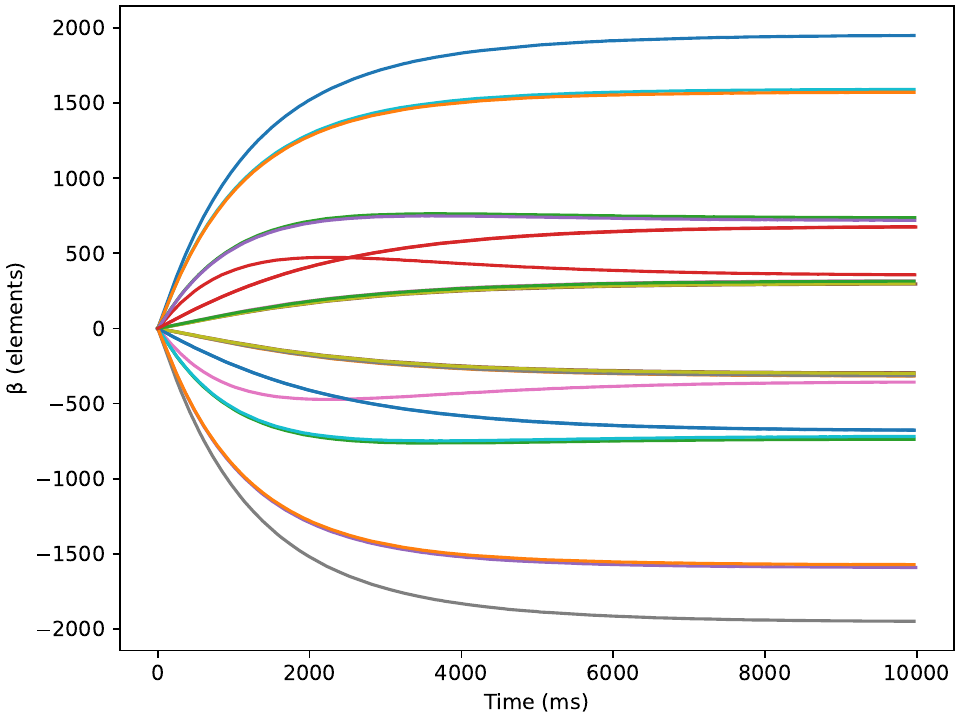}
\caption{Buffer occupancy count for the non-uniform, cube topology}
\label{fig:ebs-cube-2m}
\end{figure}

For our second experiment, we use the cube topology shown in \Cref{fig:hourglass}. 
The clock frequencies for each node can be found in \Cref{fig:clocks-cube-2m}. The buffer 
occupancy count for each node can be found in \Cref{fig:ebs-cube-2m}.

\subsection{The long link experiment}
\label{section:long-link}
\begin{figure}[htb]
  \centering
\includegraphics[width=0.9\linewidth]{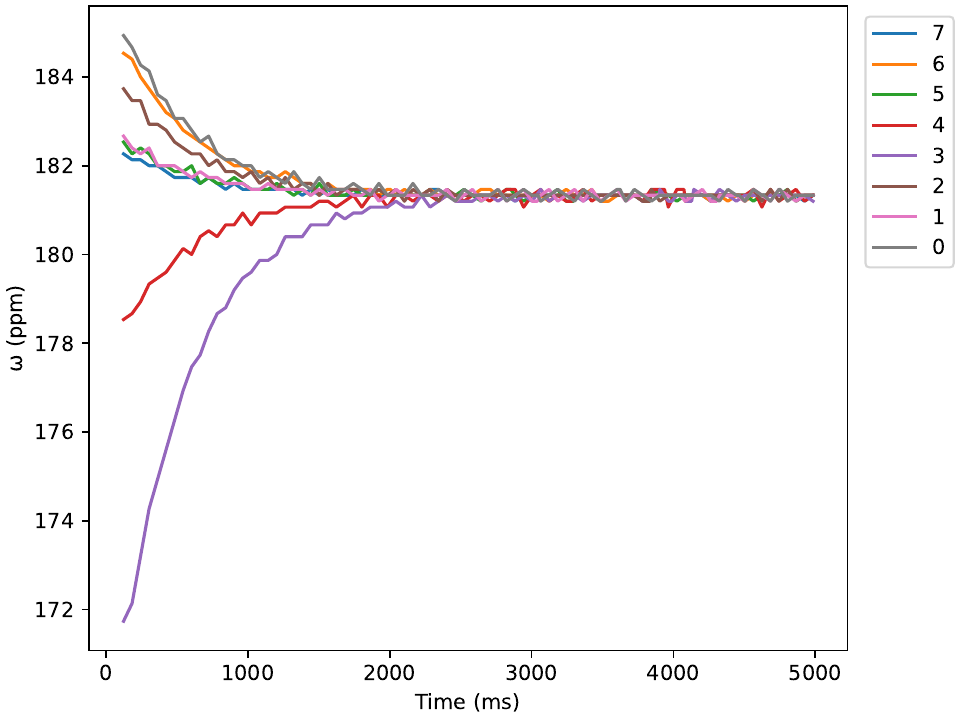}
\caption{Clock frequencies for the fully connected topology, where the link between nodes $0$ and $2$ is a \qty{2}{\kilo\meter} fiber.}
\label{fig:clocks-2km}
\end{figure}

\begin{figure}[htb]
  \centering
  \hskip -11mm\includegraphics[width=0.82\linewidth]{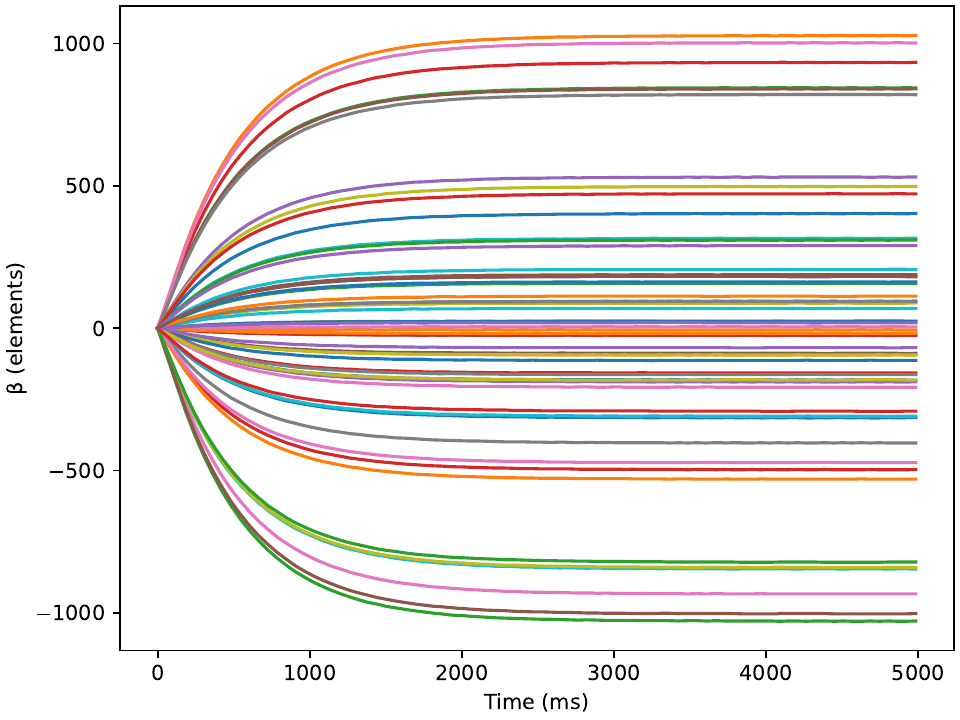}
\caption{Buffer occupancy count for the fully connected topology, where the link between nodes $0$ and $2$ is a \qty{2}{\kilo\meter} fiber.}
\label{fig:ebs-2km}
\end{figure}

For this experiment we keep all the same variables as in \Cref{subsection:fully-connected}, 
except for the link between nodes $0$~and~$2$, which is now a \qty{2}{\kilo\meter} fiber. 
The clock frequencies for each node can be found in \Cref{fig:clocks-2km}. The buffer 
occupancy count for each node can be found in \Cref{fig:ebs-2km}. The round trip logical 
latencies for each node can be found in \Cref{tab:ugns-2km}.

The reader would be forgiven for assuming this was the same data from \Cref{subsection:fully-connected}. 
We observe nearly identical clock frequencies and buffer occupancy counts, suggesting that the system 
remains insensitive to the link's physical latency, at least at this scale.

When looking a the round trip logical latencies, however, we see an
immediate difference: the replaced link jumps out with an RTT logical
latency of 1299, an increase of 1230 over its shorter
counterpart. This fits our expectations: assuming the speed of light
in a fiber is approximately two-thirds of the speed of light in a
vacuum, a increase of 1998 meter cable should increase the RTT logical
latency by 1234 --- very close to our actual measurement.

We can also use this as a proxy for estimating how many frames are in
transit within the transceivers.  First, we estimate that the full
length, \qty{2}{\kilo\meter}, should hold \mbox{$\frac{2000}{1998}1230
  = 1231$} frames. This leaves us with \mbox{$1299-1231 = 68$}
unaccounted for. Each elastic buffer is responsible for 18 frames (see
\Cref{subsection:clock-control-settings}), leaving us with \mbox{$68 -
  36 = 32$} frames, or $16$ frames per side of the transceiver.

\begin{table}[htb]
\centering
\begin{tabular}{c|ccccccc}
  FPGA & $l_1$ & $l_2$ & $l_3$ & $l_4$ & $l_5$ & $l_6$ & $l_7$ \\
  \hline
0    & 67        & 1299      & 68        & 69        & 69        & 69        & 68        \\
1    & 68        & 68        & 67        & 70        & 69        & 68        & 68        \\
2    & 68        & 1299      & 69        & 69        & 68        & 68        & 68        \\
3    & 67        & 68        & 70        & 70        & 69        & 68        & 68        \\
4    & 69        & 69        & 68        & 70        & 68        & 68        & 69        \\
5    & 68        & 70        & 67        & 69        & 69        & 68        & 69        \\
6    & 68        & 69        & 69        & 70        & 69        & 68        & 68        \\
7    & 69        & 70        & 70        & 69        & 69        & 69        & 68
\end{tabular}
\caption{Round trip logical latencies for the fully connected topology where the 
second link~$l_2$ of node~$0$ is a \qty{2}{\kilo\meter} fiber. Note that while
the link is asymmetric, the round trip logical latencies are symmetric as expected.}
\label{tab:ugns-2km}
\end{table}

\subsection{Realistic settings experiment}
\label{subsection:fast-convergence}
\begin{figure}[htb]
  \centering
  \includegraphics[width=0.9\linewidth]{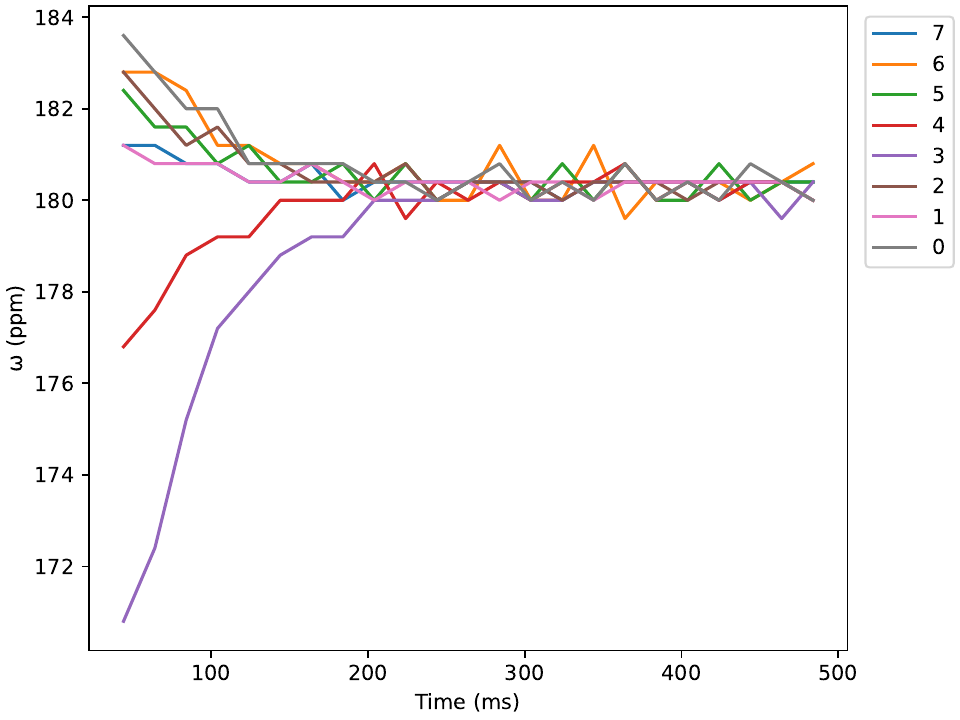}
\caption{Clock frequencies for the fully connected topology, the step size is set to \qty{0.1}{\ppm} and proportional gain to $2 \cdot 10^{-8}$. Due to the speed of convergence we are forced to sample at a higher rate, leading to more jitter in the plots.}
\label{fig:clocks-realistic}
\end{figure}

In a realistic setting we would want clock frequencies to converge as
quickly as possible, certainly faster than~\qty{50}{\second}. To this
end we would increase the step size, and pick a more aggressive
proportional gain. We set the step size to \qty{0.1}{\ppm} and the
proportional gain to $k_p = 25$.  The clock frequencies for each node
can be found in \Cref{fig:clocks-realistic}. As expected, the clocks
converge much faster: within \qty{300}{\milli\second}. At the same
time we had to increase the sampling rate to every
\qty{20}{\milli\second}, leading to more jitter in the plots.

\subsection{Measured vs calculated clock frequencies}
\begin{figure}[htb]
  \centering
  \includegraphics[width=0.9\linewidth]{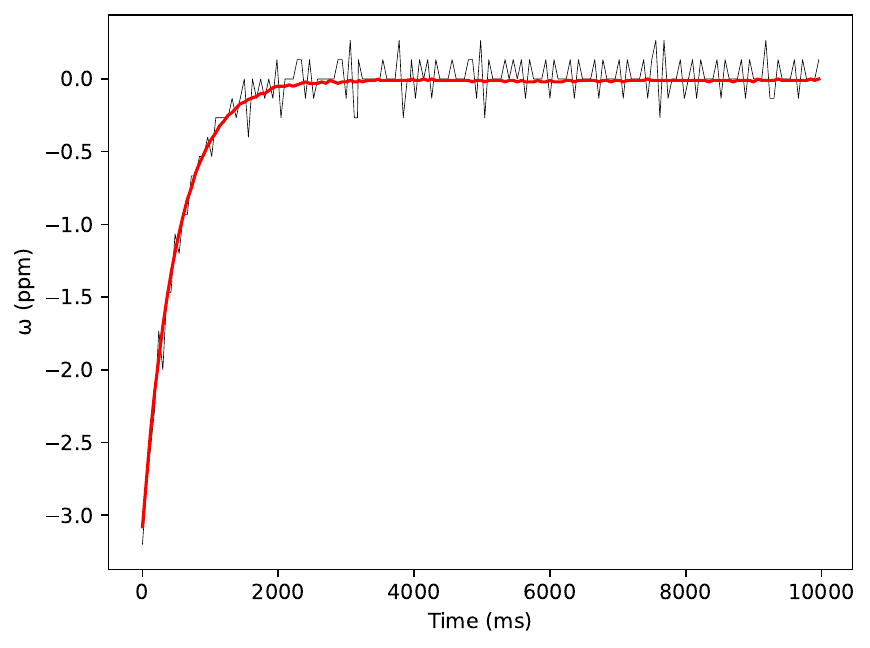}
\caption{Measured (in black) and calculated (in red) clock frequencies of one FPGA in the fully connected topology. Clock frequencies are calculated based on accumulated frequency corrections.}
\label{fig:calculated-vs-measured}
\end{figure}

Besides directly measuring the clock frequencies, we also store the accumulated
frequency corrections for each measurement. By normalizing both to zero at the
last measurement, we can compare the measured and calculated clock frequencies.
Figure \ref{fig:calculated-vs-measured} shows the results for one FPGA in the
fully connected topology. While the measured data is noisy, the calculated data
is smooth and shows the same trends. This indicates that the clock control
itself does not introduce any noise and that the noise in the measured data is
due to measurement artifacts.

\section{Mathematical modeling and validation}

A simple mathematical model for the dynamic behavior of the buffer
occupancies in the \bt mechanism has been developed
in~\cite{bms}. This model is called the \emph{abstract~frame~model}.
\begin{align*}
  \frac{d\theta_i}{dt} &= \omega_i(t) \\
  \beta_\jtoi(t) &= \floor{\theta_j(t-l_\jtoi)} - \floor{\theta_i(t)} + \ugn_\jtoi \\
  \omega_i(t) &=   \omega_i^k \quad  \text{for } t \in [s_i^k, s_i^{k+1})\\
    \theta_i(t_i^k) &= \theta_i^0 + kp \\
    \theta_i(s_i^k) &= \theta^0_i + kp+d  \\
    \crel_i(t_i^k) &= k_p \sum_{j \mid \jtoi} (\beta_\jtoi(t_i^k) - \beta_i^\text{off})\\
    \omega_i^k & = \wu_i ( 1+  \crel_i(t_i^k))
\end{align*}
Here $\omega_i(t)$ is the frequency of the clock at node $i$ at time
$t$, and $\theta_i(t)$ is the corresponding clock phase, measured in
units of localticks. Specifically, a localtick is defined to occur
whenever $\theta_i$ is an integer. The physical latency of the link $j
\to i$ is $l_\jtoi$, and the logical latency is $\lambda_\jtoi$.  At
node $i$, there is an elastic buffer for each incoming link $j \to i$
with buffer occupancy $\beta_\jtoi$.  The buffer occupancies are
measured periodically with period $p$ localticks with respect to the
clock at node $i$. There is a delay of $d$ localticks between the
measurement of the buffer occupancy and updating the oscillator
frequency. In the above model, $t_i^k$ is the physical time at which
measurements occur and $s_i^k$ is the physical time at which
oscillator frequency updates occur. The controller sets the 
oscillator correction $\crel(t_i^k)$ for node $i$ at time step $k$.
The clock phase at startup is $\theta_i^0$.

We know almost all of the parameters for this system. There are some
parameters which are known only to limited accuracy, in particular the
physical latencies $l$ and the oscillator update delays $d$. However,
the behavior of the system is insensitive to both of these parameters.
In particular, the lack of latency sensitivity is illustrated
by the experiments above where a 2km fiber is used. Not all of the parameters
may be chosen independently; for example, the logical latencies $\lambda$
are determined by the initial clock phases and initial buffer occupancies.

We simulate the system with the hourglass topology, and compare with
the data measured in the experiment.  We used the open-source
Callisto~\cite{callisto} simulator. The unknown frequencies $\wu$ for
the model are determined by the initial frequencies observed in the
hardware, which are at the left-hand edge of
Figure~\ref{fig:clocks-hourglass-2m}.  We can see that the frequency
dynamics of the simulation match closely with the experiment.

\begin{figure}[htb]
  \centerline{\begin{overpic}
      [scale=0.07]{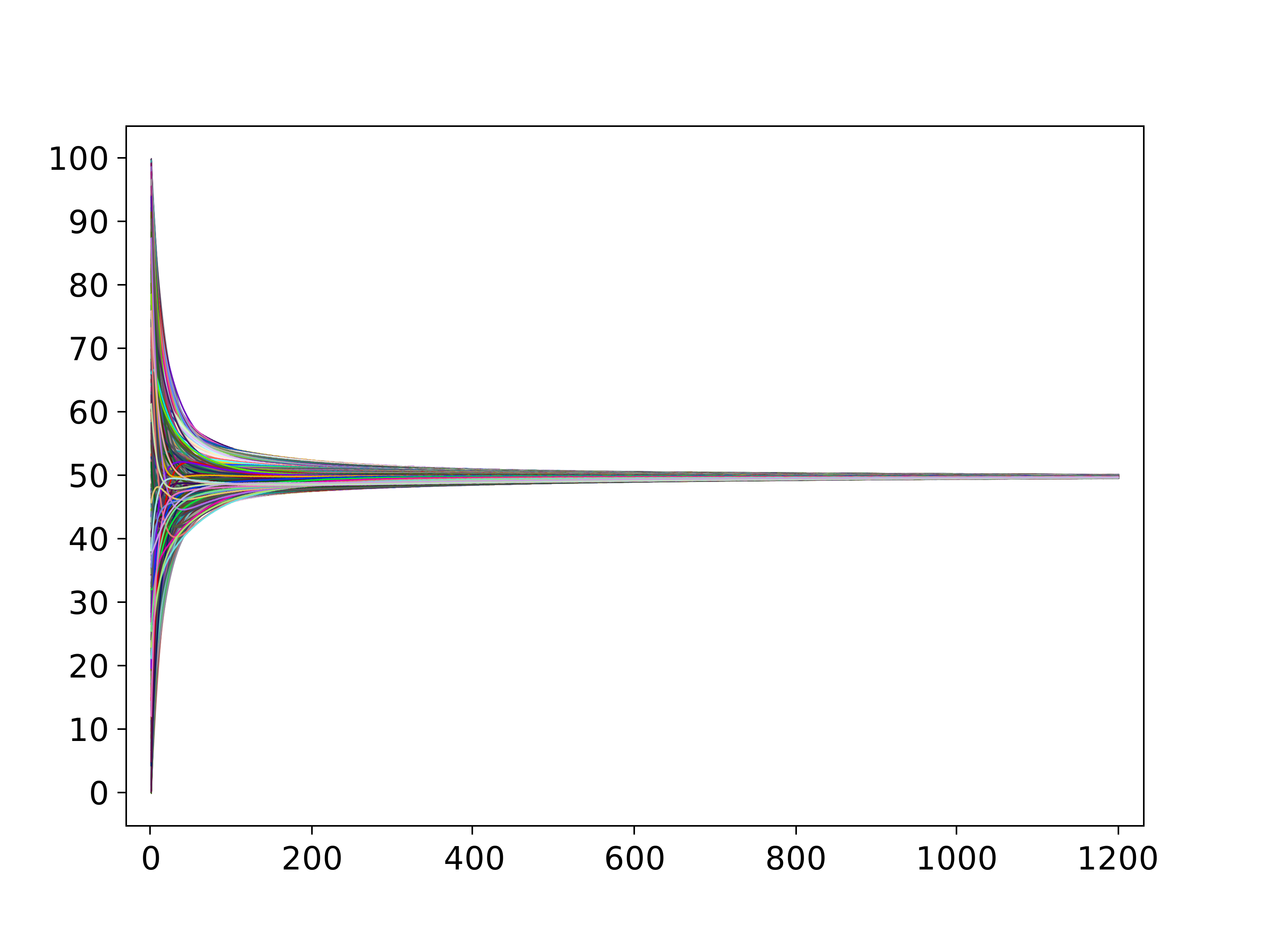}
      \put(50,1){\clap{\small\text{Time(ms)}}}
      \put(0,36){\clap{\vcent{\rotatebox{90}{\small $\omega$ (ppm)}}}}
  \end{overpic}}
  \vskip1mm
  \caption{Clock frequencies for a 3d torus with $22^3$ nodes.}
  \label{fig:large}
\end{figure}

There are several purposes of this modeling effort. One of the most
important is that, in order to design the frequency control algorithm
inside the \bt mechanism, one must select between many possible
alternatives. Being able to reproduce reliably the behavior of the
system using a simple model allows for fast simulation, as well as
affording the possibility of using mathematical tools to predict
performance; an example of such prediction is given in~\cite{res}. We
would also like to understand how the system scales, \eg, how long
does it takes for buffer occupancies to converge when there are many
thousands of nodes.  This can be easily and accurately determined in
simulation. An example is shown in Figure~\ref{fig:large} of a system
with $22^3$ nodes arranged in a 3-dimensional torus. This simulation
was performed in Callisto, and shows that for such large networks
we see the expected convergence of frequencies.

\section{Related work}
\label{section:relatedwork}

Achieving synchrony in distributed systems is a long-standing challenge.  
Existing approaches can be broadly categorized by their level of 
synchronization and reliance on a global time reference.

\paragraph{Clock synchronization and syntonization.} Traditional methods, like 
Network Time Protocol (NTP)~\cite{ptp}, aim to align local clocks to a 
global reference, often a UTC server. More specialized systems, 
like SONET~\cite{sonet} in telecom and SyncE~\cite{syncE} in networking 
equipment, strive for clock syntonization, ensuring all reference 
oscillators maintain the same average frequency. These approaches 
typically rely on hierarchical clock distribution, deriving timing 
from upstream communication links and aligning to a global master clock.

\paragraph{Logical synchrony and \bt.} In contrast to global clock 
synchronization, logical synchrony~\cite{ls} offers a framework for 
synchronous distributed computation without relying on a global clock. 
The \bt mechanism, introduced in this paper~\cite{bms}, leverages 
logical synchrony, enabling precise coordination in a distributed 
system through decentralized frequency alignment and local elastic buffers.
This approach eliminates the need for hierarchical clock distribution 
and complex handshakes, resulting in a scalable and efficient 
synchronization mechanism.

\paragraph{Theoretical foundations.} The theoretical underpinnings of \bt, 
including its mathematical model, have been explored in previous 
works~\cite{spalink_2006, bms, res}. Notably, the model shares similarities 
with extensively studied linear consensus models found in diverse applications 
like flocking, averaging models, and congestion control~\cite{dorfler, 
moreau2004, boyds, hastings, boyd2004, kelly, 
strogatz, bullo, hedrick, olfati, slotine, burbano, freeman, andreasson}.

\begin{figure}[htb]
  \centerline{\begin{overpic}
      [scale=0.07]{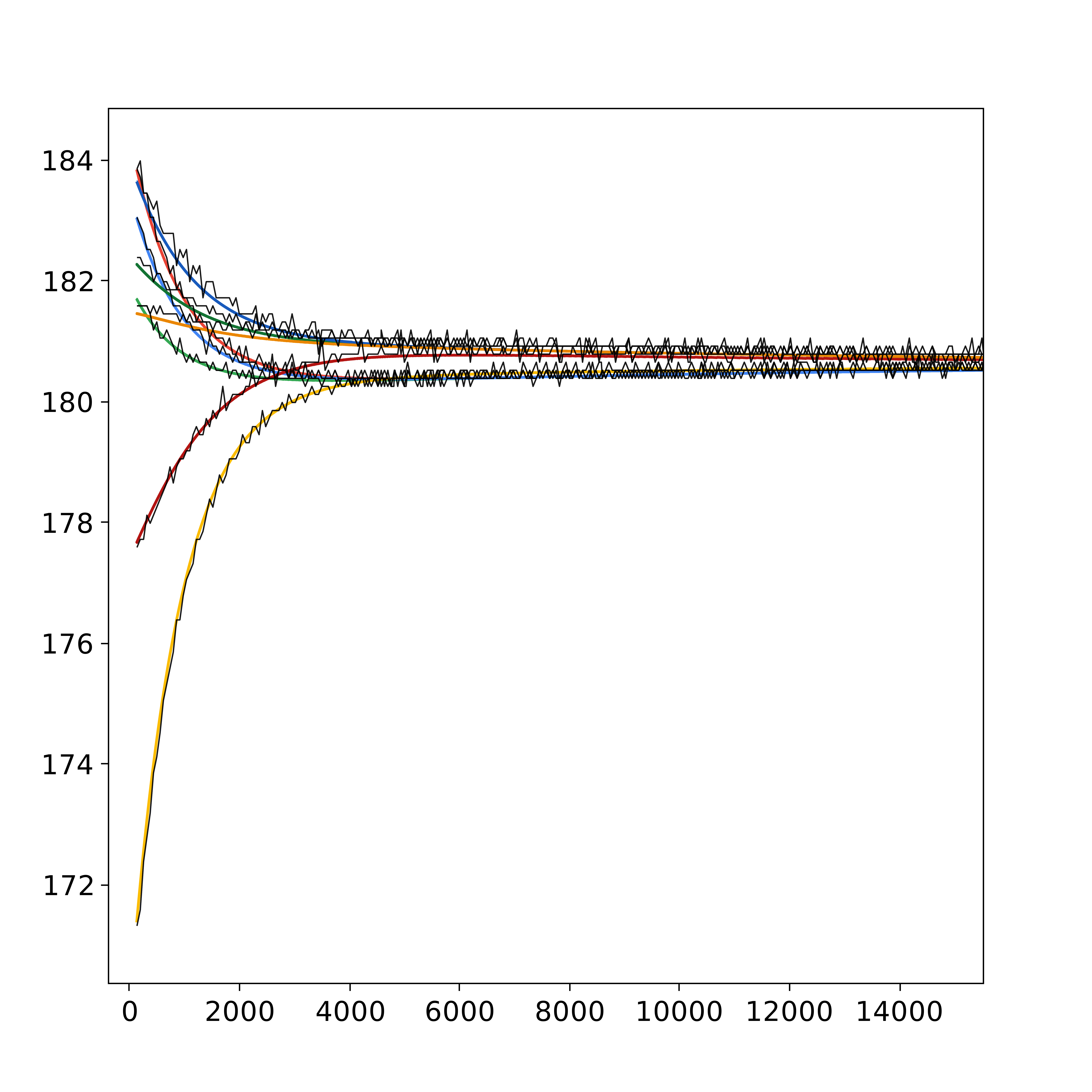}
      \put(50,1){\clap{\small\text{Time(ms)}}}
      \put(0,50){\clap{\vcent{\rotatebox{90}{\small $\omega$ (ppm)}}}}
  \end{overpic}}
  \vskip1mm
  \caption{Comparison of clock frequencies for the hourglass topology. The black
    lines are measured from the FPGA,
    the colored lines are generated by a simulation of the mathematical model.}
  \label{fig:corr}
\end{figure}

\paragraph{Synchronous execution and programming models.} Lamport's seminal 
work~\cite{lamport} established the concept of consistent event ordering 
and global logical clocks in distributed systems, leading to the development 
of vector clocks~\cite{mattern, fidge} for capturing this ordering at each node. 
This paved the way for synchronous execution 
models~\cite{lamport_reliable_distr_systems_1978, dwork_consensus_1988, 
liskov_practical_1991, le2003polychrony, lee:time:cacm:2009}, particularly 
relevant for cyber-physical systems, where a global time reference simplifies 
reasoning, correctness proofs, and algorithm design.  

Various synchronous programming languages and models, such as 
Esterel~\cite{berry:esterel:1998}, Lustre~\cite{halbwachs:lustre:1991}, 
Signal~\cite{benveniste:signal:1991}, Synchronous Dataflow~\cite{lee:sdf:1987}, GALS 
communication models~\cite{potop-butucaru_concurrency_2004}, 
and Timed CSP~\cite{reed1988metric}, have emerged, each addressing specific 
aspects and applications of synchronous computation.  These developments 
demonstrate the diverse landscape of synchronous models and highlight the 
importance of choosing the right abstraction for a given distributed 
system~\cite{benveniste:12yearsofsynchrony:2003, didier_sheep_2019}.

\section{Conclusions}

This work presents the first hardware implementation of the \bt synchronization 
mechanism, demonstrating its ability to achieve network-wide frequency alignment 
and establish a consistent notion of logical time.  Through experiments on an 8-node 
network of FPGA boards, we validated that \bt effectively synchronizes clocks 
across a range of network topologies, including fully connected, hourglass, and 
cube configurations, even in the presence of varying physical link latencies.
Our hardware implementation successfully converges to a stable state where all 
nodes operate continuously within sufficiently tight frequency bounds to avoid 
any buffer over- or under-runs.  We also verified that the logical latencies in 
the system remain stable and predictable, even when a long 2km fiber link is introduced. 
This robust behavior confirms the effectiveness of the decentralized clock control 
mechanism and the insensitivity of the system to physical latency variations, aligning 
with theoretical predictions and simulations.
 
\bt is a fundamentally deterministic network design which allows scheduling of compute 
and communication together, which in turn allows a level of efficiency otherwise impossible.
Techniques like dynamic flow control and wall clock synchronization impose nondeterminism
and overhead costs.  Our solution allows these to be eliminated for performance critical 
applications with predictable resource needs.

This work concretely shows that \bt is implementable, and therefore suggests it should 
be considered as a practical, scalable, and efficient synchronization solution for 
distributed systems. The hardware implementation lays the foundation for future research 
exploring \bt performance at larger scales, its application in real-world distributed 
computing scenarios, and the development of higher-level programming abstractions 
built upon its robust logical synchrony guarantees.

\def\makeack{
\section{Acknowledgments}
The research in this paper came about through a great deal of
collaborative effort.  In particular, we would like to thank Lucas
Bollen, Lara Herzog, Hidde Moll, and Leon Schoorl for their extensive
work on the \bt implementation. We thank C\u{a}lin Ca\c{s}caval for
his careful review of the draft manuscript and for a longstanding and
fruitful collaboration.
}

\if\MODE1\makeack\fi

\input{main.bbl}
\end{document}

%% file: preamble.tex
\usepackage{stmaryrd}
\usepackage{cleveref}
\usepackage{tabularx}
\usepackage{siunitx}
\usepackage{wasysym}
\usepackage{fancyvrb}
\usepackage{overpic}
\usepackage{enumitem}
\usepackage{mathtools}
\usepackage{xspace}
\usepackage{comment}
\usepackage[margin=10pt,skip=7pt,font=small]{caption}

\definecolor{tred}{rgb}{0.858, 0.888, 0.978}
\graphicspath{{figures/}}

\usepackage[textsize=small,backgroundcolor=tred]{todonotes}

\usepackage[numbers,sort]{natbib}

\usepackage{etoolbox}
\apptocmd{\sloppy}{\hbadness 10000\relax}{}{}

\DeclareSIUnit\ppb{ppb}
\DeclareSIUnit\ppm{ppm}

\newcommand{\syncsym}{\rlap{\raise0.08em\hbox{\kern0.25em\scalebox{0.6}{\texttt{R}}}}$\circlearrowright$}
\newcommand{\fdetect}{$\urcorner$\kern-0.16em$\llcorner$}

\newcommand{\from}{\ensuremath{\colon}}
\renewcommand{\to}{\ensuremath{\rightarrow}}
\newcommand{\bools}{\ensuremath{\mathbb{B}}}
\newcommand{\nats}{\ensuremath{\mathbb{N}}}
\newcommand{\ints}{\ensuremath{\mathbb{Z}}}

\newcommand{\concat}{\ensuremath{\diamond}}

\newbox\vcbox
\def\vcent#1{\setbox\vcbox\hbox{#1}\raise -0.5\ht\vcbox\hbox{#1}}

\def\wu{\omega^\text{u}}

\def\crel{c^\text{rel}}
\def\cest{c^\text{est}}
\def\cinc{c^\text{inc}}
\def\finc{\texttt{FINC}\xspace}
\def\fdec{\texttt{FDEC}\xspace}

\usepackage{stmaryrd}
\usepackage{trimclip}
\makeatletter
\DeclareRobustCommand{\shortto}{\mathrel{\mathpalette\short@to\relax}}
\newcommand{\short@to}[2]{\mkern2mu
  \clipbox{{.5\width} 0 0 0}{$\m@th#1\vphantom{+}{\shortrightarrow}$}}
\makeatother
\def\itoj{{i \shortto j}}
\def\jtoi{{j \shortto i}}

\newcommand{\eg}{{\it e.g.}}
\newcommand{\ie}{{\it i.e.}}
\def\ugn{\lambda}

\DeclarePairedDelimiter{\floor}{\lfloor}{\rfloor}
\newcommand{\bt}{bittide\xspace}

\font\fivefont=cmss10 at 5pt
\def\doublefont{\fontencoding{T1}\fontfamily{cmss}\fontsize{11pt}{13}\selectfont}
\def\fivept{\fivefont}

%% file: main.bbl
\begin{thebibliography}{50}
\providecommand{\natexlab}[1]{#1}
\providecommand{\url}[1]{\texttt{#1}}
\expandafter\ifx\csname urlstyle\endcsname\relax
  \providecommand{\doi}[1]{doi: #1}\else
  \providecommand{\doi}{doi: \begingroup \urlstyle{rm}\Url}\fi

\bibitem[Taft et~al.(2020)Taft, Sharif, Matei, {VanBenschoten}, Lewis, Grieger,
  Niemi, Woods, Birzin, Poss, Bardea, Ranade, Darnell, Gruneir, Jaffray, Zhang,
  and Mattis]{cockroachdb}
R.~Taft, I.~Sharif, A.~Matei, N.~{VanBenschoten}, J.~Lewis, T.~Grieger,
  K.~Niemi, A.~Woods, A.~Birzin, R.~Poss, P.~Bardea, A.~Ranade, B.~Darnell,
  B.~Gruneir, J.~Jaffray, L.~Zhang, and P.~Mattis.
\newblock {CockroachDB: The} resilient geo-distributed {SQL} database.
\newblock In \emph{Proceedings of the 2020 ACM SIGMOD international conference
  on management of data}, pages 1493--1509, 2020.

\bibitem[Li et~al.(2020)Li, Kumar, Hariharan, Wassel, Hochschild, Platt,
  Sabato, Yu, Dukkipati, Chandra, and Vahdat]{li_sundial_2020}
Y.~Li, G.~Kumar, H.~Hariharan, H.~Wassel, P.~Hochschild, D.~Platt, S.~Sabato,
  M.~Yu, N.~Dukkipati, P.~Chandra, and A.~Vahdat.
\newblock Sundial: {Fault-tolerant} clock synchronization for datacenters.
\newblock In \emph{14th {USENIX} Symposium on Operating Systems Design and
  Implementation ({OSDI} 20)}, pages 1171--1186, November 2020.

\bibitem[Corbett et~al.(2013)Corbett, Dean, Epstein, Fikes, Frost, Furman,
  Ghemawat, Gubarev, Heiser, Hochschild, Hsieh, Kanthak, Kogan, Li, Lloyd,
  Melnik, Mwaura, Nagle, Quinlan, Rao, Rolig, Saito, Szymaniak, Taylor, Wang,
  and Woodford]{corbett_spanner_2013}
J.~C. Corbett, J.~Dean, M.~Epstein, Andrew Fikes, C.~Frost, J.~J. Furman,
  S.~Ghemawat, A.~Gubarev, C.~Heiser, P.~Hochschild, W.~Hsieh, S.~Kanthak,
  E.~Kogan, H.~Li, A.~Lloyd, S.~Melnik, D.~Mwaura, D.~Nagle, S.~Quinlan,
  R.~Rao, L.~Rolig, Y.~Saito, M.~Szymaniak, C.~Taylor, R.~Wang, and
  D.~Woodford.
\newblock Spanner: {Google}’s globally distributed database.
\newblock \emph{ACM Transactions on Computer Systems}, 31\penalty0
  (3):\penalty0 1--22, 2013.

\bibitem[Gupta et~al.(2023)Gupta, Goyal, Marinos, Zhao, Mittal, and
  Chandra]{gupta}
E.~Gupta, P.~Goyal, I.~Marinos, C.~Zhao, R.~Mittal, and R.~Chandra.
\newblock {DBO: Fairness} for cloud-hosted financial exchanges.
\newblock In \emph{Proceedings of ACM SIGCOMM}, pages 550--563, 2023.

\bibitem[Bateni et~al.(2022)Bateni, Lohstroh, Wong, Tabish, Kim, Lin, Menard,
  Liu, and Lee]{bateni}
S.~Bateni, M.~Lohstroh, H.~S. Wong, R.~Tabish, H.~Kim, S.~Lin, C.~Menard,
  C.~Liu, and E.~A. Lee.
\newblock {Xronos: Predictable} coordination for safety-critical distributed
  embedded systems, 2022.
\newblock \url{https://arxiv.org/abs/2207.09555}.

\bibitem[Cowan et~al.(2023)Cowan, Maleki, Musuvathi, Saarikivi, and
  Xiong]{cowan}
M.~Cowan, S.~Maleki, M.~Musuvathi, O.~Saarikivi, and Y.~Xiong.
\newblock {MSCCLang: Microsoft} collective communication language.
\newblock In \emph{ACM International Conference on Architectural Support for
  Programming Languages and Operating Systems}, pages 502--514, 2023.

\bibitem[Lall et~al.(2024)Lall, Ca\c{s}caval, Izzard, and Spalink]{ls}
S.~Lall, C.~Ca\c{s}caval, M.~Izzard, and T.~Spalink.
\newblock Logical synchrony and the bittide mechanism.
\newblock \emph{IEEE Transactions on Parallel and Distributed Systems},
  35\penalty0 (11):\penalty0 1936--1948, November 2024.

\bibitem[ptp()]{ptp}
Precision clock synchronization protocol for networked measurement and control
  systems.
\newblock IEEE standard 2021.9456762.

\bibitem[Spalink(2006)]{spalink_2006}
T.~Spalink.
\newblock \emph{Deterministic sharing of distributed resources}.
\newblock Princeton University, 2006.

\bibitem[Lall et~al.(2022{\natexlab{a}})Lall, Ca\c{s}caval, Izzard, and
  Spalink]{bms}
S.~Lall, C.~Ca\c{s}caval, M.~Izzard, and T.~Spalink.
\newblock Modeling and control of bittide synchronization.
\newblock In \emph{Proceedings of the American Control Conference}, pages
  5185--5192, 2022{\natexlab{a}}.
\newblock Available at \url{https://arxiv.org/abs/2109.14111}.

\bibitem[Lamport(1978{\natexlab{a}})]{lamport}
L.~Lamport.
\newblock Time, clocks, and the ordering of events in a distributed system.
\newblock \emph{Communications of the ACM}, 21\penalty0 (7):\penalty0
  558–565, 1978{\natexlab{a}}.

\bibitem[cal()]{callisto}
Callisto: {Simulator} of bittide clock synchronization dynamics.
\newblock \url{https://github.com/bittide/ Callisto.jl}.

\bibitem[aeg()]{aegir}
Aegir: {Multi-level} bittide functional simulator.
\newblock \url{https://github.com/bittide/aegir}.

\bibitem[qba()]{qbay}
The bittide-hardware implementation of the bittide system.
\newblock \url{https://github.com/bittide/ bittide-hardware}.

\bibitem[Lall et~al.(2023)Lall, Ca\c{s}caval, Izzard, and Spalink]{reframing}
S.~Lall, C.~Ca\c{s}caval, M.~Izzard, and T.~Spalink.
\newblock On buffer centering for bittide synchronization.
\newblock In \emph{International Conference on Control, Decision, and
  Information Technologies}, 2023.
\newblock Available at \url{https://arxiv.org/abs/2303.11467}.

\bibitem[{Telcordia GR-253}(2000)]{sonet}
{Telcordia GR-253}.
\newblock Synchronous {O}ptical {N}etwork ({SONET}) {T}ransport {S}ystems:
  {C}ommon {G}eneric {C}riteria, 2000.

\bibitem[SyncE()]{syncE}
SyncE.
\newblock {T}iming and {S}ynchronization {A}spects in {P}acket {N}etworks.
\newblock ITU-T G.8261/Y.1361.

\bibitem[Lipinski et~al.(2011)Lipinski, Wlostowski, Serrano, and
  Alvarez]{whiterabbit}
M.~Lipinski, T.~Wlostowski, J.~Serrano, and P.~Alvarez.
\newblock White rabbit: a {PTP} application for robust sub-nanosecond
  synchronization.
\newblock In \emph{IEEE International Symposium on Precision Clock
  Synchronization for Measurement, Control and Communication}, pages 25--30,
  2011.

\bibitem[Lall and Spalink(2025)]{mbo}
S.~Lall and T.~Spalink.
\newblock Modeling buffer occupancy in bittide systems.
\newblock To appear, American Control Conference, 2025.

\bibitem[Gray(1953)]{gray}
F.~Gray.
\newblock Pulse code communication.
\newblock \emph{United States Patent Number 2632058}, 1953.
\newblock \url{https://cir.nii.ac.jp/crid/1572261550584107136}.

\bibitem[Lall et~al.(2022{\natexlab{b}})Lall, Ca\c{s}caval, Izzard, and
  Spalink]{res}
S.~Lall, C.~Ca\c{s}caval, M.~Izzard, and T.~Spalink.
\newblock Resistance distance and control performance for bittide
  synchronization.
\newblock In \emph{Proceedings of the European Control Conference}, pages
  1850--1857, 2022{\natexlab{b}}.
\newblock Available at \url{https://arxiv.org/abs/2111.05296}.

\bibitem[Dorfler and Bullo(2014)]{dorfler}
F.~Dorfler and F.~Bullo.
\newblock {S}ynchronization in {C}omplex {N}etworks of {P}hase {O}scillators:
  {A} {S}urvey.
\newblock \emph{Automatica}, 50\penalty0 (6):\penalty0 1539--1564, 2014.

\bibitem[Moreau(2004)]{moreau2004}
L.~Moreau.
\newblock {S}tability of {C}ontinuous-time {D}istributed {C}onsensus
  {A}lgorithms.
\newblock In \emph{Proceedings of the IEEE Conference on Decision and Control},
  volume~4, pages 3998--4003, 2004.

\bibitem[Reynolds(1987)]{boyds}
C.~W. Reynolds.
\newblock Flocks, herds and schools: a distributed behavioral model.
\newblock In \emph{Proceedings 14th ACM SIGGRAPH}, pages 25--34, 1987.

\bibitem[Hastings(1970)]{hastings}
W.~K. Hastings.
\newblock \emph{Monte Carlo sampling methods using {Markov} chains and their
  applications}.
\newblock Oxford University Press, 1970.

\bibitem[Boyd et~al.(2004)Boyd, Diaconis, and Xiao]{boyd2004}
S.~P. Boyd, P.~Diaconis, and L.~Xiao.
\newblock Fastest mixing {Markov} chain on a graph.
\newblock \emph{SIAM Review}, 46\penalty0 (4):\penalty0 667--689, 2004.

\bibitem[Kelly et~al.(1998)Kelly, Maulloo, and Tan]{kelly}
F.~P. Kelly, A.~K. Maulloo, and D.~Tan.
\newblock Rate control for communication networks: shadow prices, proportional
  fairness and stability.
\newblock \emph{Journal of the Operational Research Society}, 49\penalty0
  (3):\penalty0 237--252, 1998.

\bibitem[Strogatz(2000)]{strogatz}
S.~Strogatz.
\newblock From {Kuramoto} to {Crawford}: exploring the onset of synchronization
  in populations of coupled oscillators.
\newblock \emph{Physica D: Nonlinear Phenomena}, 143\penalty0 (1-4):\penalty0
  1--20, 2000.

\bibitem[Dorfler and Bullo(2012)]{bullo}
F.~Dorfler and F.~Bullo.
\newblock Synchronization and transient stability in power networks and
  nonuniform kuramoto oscillators.
\newblock \emph{SIAM Journal on Control and Optimization}, 50\penalty0
  (3):\penalty0 1616--1642, 2012.
\newblock \doi{10.1137/110851584}.

\bibitem[Swaroop and Hedrick(1996)]{hedrick}
D.~Swaroop and J.~K. Hedrick.
\newblock String stability of interconnected systems.
\newblock \emph{IEEE Transactions on Automatic Control}, 41\penalty0
  (3):\penalty0 349--357, 1996.

\bibitem[Olfati-Saber and Murray(2004)]{olfati}
R.~Olfati-Saber and R.M. Murray.
\newblock Consensus problems in networks of agents with switching topology and
  time-delays.
\newblock \emph{IEEE Transactions on Automatic Control}, 49\penalty0
  (9):\penalty0 1520--1533, 2004.
\newblock \doi{10.1109/TAC.2004.834113}.

\bibitem[Wang and Slotine(2005)]{slotine}
W.~Wang and J.-J.~E. Slotine.
\newblock On partial contraction analysis for coupled nonlinear oscillators.
\newblock \emph{Biological Cybernetics}, 92\penalty0 (1):\penalty0 38--53,
  2005.

\bibitem[Burbano~Lombana and di~Bernardo(2015)]{burbano}
D.~Burbano~Lombana and M.~di~Bernardo.
\newblock Distributed {PID} control for consensus of homogeneous and
  heterogeneous networks.
\newblock \emph{IEEE Transactions on Control of Network Systems}, 2\penalty0
  (2):\penalty0 154--163, 2015.

\bibitem[Freeman et~al.(2006)Freeman, Yang, and Lynch]{freeman}
R.~A. Freeman, P.~Yang, and K.~M. Lynch.
\newblock Stability and convergence properties of dynamic average consensus
  estimators.
\newblock In \emph{Proceedings of the 45th IEEE Conference on Decision and
  Control}, pages 338--343, 2006.

\bibitem[Andreasson et~al.(2014)Andreasson, Dimarogonas, Sandberg, and
  Johansson]{andreasson}
M.~Andreasson, D.~V. Dimarogonas, H.~Sandberg, and K.~H. Johansson.
\newblock Distributed control of networked dynamical systems: Static feedback,
  integral action and consensus.
\newblock \emph{IEEE Transactions on Automatic Control}, 59\penalty0
  (7):\penalty0 1750--1764, 2014.

\bibitem[Mattern(1989)]{mattern}
F.~Mattern.
\newblock Virtual time and global states of distributed systems.
\newblock In \emph{Proceedings of the workshop on parallel and distributed
  algorithms}, pages 215--226, 1989.

\bibitem[Fidge(1988)]{fidge}
C.~J. Fidge.
\newblock Timestamps in message-passing systems that preserve the partial
  ordering.
\newblock \emph{Australian Computer Science Communications}, 10\penalty0
  (1):\penalty0 56--66, February 1988.

\bibitem[Lamport(1978{\natexlab{b}})]{lamport_reliable_distr_systems_1978}
Leslie Lamport.
\newblock The implementation of reliable distributed multiprocess systems.
\newblock \emph{Computer Networks}, 2\penalty0 (2):\penalty0 95--114,
  1978{\natexlab{b}}.

\bibitem[Dwork et~al.(1988)Dwork, Lynch, and Stockmeyer]{dwork_consensus_1988}
C.~Dwork, N.~Lynch, and L.~Stockmeyer.
\newblock Consensus in the presence of partial synchrony.
\newblock \emph{Journal of the ACM}, 35\penalty0 (2):\penalty0 288--323, 1988.

\bibitem[Liskov(1991)]{liskov_practical_1991}
B.~Liskov.
\newblock Practical uses of synchronized clocks in distributed systems.
\newblock In \emph{Proceedings of the {ACM} symposium on principles of
  distributed computing}, pages 1--9, 1991.

\bibitem[Le~Guernic et~al.(2003)Le~Guernic, Talpin, and
  Le~Lann]{le2003polychrony}
P.~Le~Guernic, J.-P. Talpin, and J.-C. Le~Lann.
\newblock Polychrony for system design.
\newblock \emph{Journal of Circuits, Systems, and Computers}, 12\penalty0
  (03):\penalty0 261--303, 2003.

\bibitem[Lee(2009)]{lee:time:cacm:2009}
E.~A. Lee.
\newblock Computing needs time.
\newblock \emph{Communications of the ACM}, 52\penalty0 (5):\penalty0 70--79,
  2009.

\bibitem[Berry(1998)]{berry:esterel:1998}
G.~Berry.
\newblock The {Foundations} of {Esterel}, 1998.

\bibitem[Halbwachs et~al.(1991)Halbwachs, Caspi, Raymond, and
  Pilaud]{halbwachs:lustre:1991}
N.~Halbwachs, P.~Caspi, P.~Raymond, and D.~Pilaud.
\newblock The synchronous data flow programming language {LUSTRE}.
\newblock \emph{Proceedings of the IEEE}, 79\penalty0 (9):\penalty0 1305--1320,
  1991.

\bibitem[Benveniste et~al.(1991)Benveniste, Guernic, and
  Jacquemot]{benveniste:signal:1991}
A.~Benveniste, P.~Le Guernic, and C.~Jacquemot.
\newblock Synchronous programming with events and relations: the {SIGNAL}
  language and its semantics.
\newblock \emph{Science of Computer Programming}, 16\penalty0 (2):\penalty0
  103--149, 1991.

\bibitem[Lee and Messerschmitt(1987)]{lee:sdf:1987}
E.~A. Lee and D.~G. Messerschmitt.
\newblock Static scheduling of synchronous data flow programs for digital
  signal processing.
\newblock \emph{IEEE Transactions on Computers}, C-36\penalty0 (1):\penalty0
  24--35, 1987.

\bibitem[Potop-Butucaru et~al.(2004)Potop-Butucaru, Caillaud, and
  Benveniste]{potop-butucaru_concurrency_2004}
D.~Potop-Butucaru, B.~Caillaud, and A.~Benveniste.
\newblock Concurrency in synchronous systems.
\newblock In \emph{International conference on application of concurrency to
  system design}, pages 67--76, 2004.

\bibitem[Reed and Roscoe(1988)]{reed1988metric}
G.~M. Reed and A.~W. Roscoe.
\newblock Metric spaces as models for real-time concurrency.
\newblock In \emph{Workshop on mathematical foundations of programming language
  semantics}, pages 331--343, 1988.

\bibitem[Benveniste et~al.(2003)Benveniste, Caspi, Edwards, Halbwachs,
  Le~Guernic, and de~Simone]{benveniste:12yearsofsynchrony:2003}
A.~Benveniste, P.~Caspi, S.A. Edwards, N.~Halbwachs, P.~Le~Guernic, and
  R.~de~Simone.
\newblock The synchronous languages 12 years later.
\newblock \emph{Proceedings of the IEEE}, 91\penalty0 (1):\penalty0 64--83,
  2003.

\bibitem[Didier et~al.(2019)Didier, Cohen, Potop-Butucaru, and
  Gauffriau]{didier_sheep_2019}
K.~Didier, A.~Cohen, D.~Potop-Butucaru, and A.~Gauffriau.
\newblock Sheep in wolf's clothing: {Implementation} models for dataflow
  multi-threaded software.
\newblock In \emph{International Conference on Application of Concurrency to
  System Design}, pages 43--52, June 2019.

\end{thebibliography}
